 \documentclass[prd,aps,preprintnumbers,amsmath,amssymb,nofootinbib,eqsecnum, preprintnumbers]{revtex4}
 \pdfoutput=1
\usepackage{graphicx}
\usepackage{epsf}
\usepackage{amsmath}
\usepackage{epstopdf}
\usepackage{bm}
\usepackage{color}
\usepackage{tabularx}
\usepackage{enumitem}
\usepackage{float}
\usepackage{array,booktabs}
\usepackage{footnote}
\usepackage{threeparttable}
\usepackage{graphicx}
\usepackage{hyperref}
\usepackage{amssymb,epsf}
\usepackage{latexsym}
\usepackage{epstopdf}
\usepackage{epsfig}
\begin{document}


\title{Thermodynamic schemes of charged BTZ-like black holes in arbitrary dimensions}


\author{Ali Dehghani,$^{1,}$\footnote{email address: ali.dehghani.phys@gmail.com}
Behnam Pourhassan,$^{1,}$\footnote{email address: b.pourhassan@du.ac.ir} Soodeh 
Zarepour,$^{2,}$\footnote{email address: szarepour@phys.usb.ac.ir}
and Emmanuel 
N. Saridakis,$^{3,4,5,}$\footnote{email address: msaridak@noa.gr}
}

\affiliation{$^1$School of Physics, Damghan University, Damghan 3671645667, Iran}

\affiliation{$^2$Department of Physics, University of Sistan and Baluchestan, Zahedan, Iran}

\affiliation{$^3$National Observatory of Athens, Lofos Nymfon, 11852 Athens, 
Greece}

\affiliation{$^4$CAS Key Laboratory for Researches in Galaxies and Cosmology, 
Department of Astronomy, University of Science and Technology of China, Hefei, 
Anhui 230026, P.R. China}
 \affiliation{$^5$Departamento de Matem\'{a}ticas, Universidad Cat\'{o}lica del 
Norte,  Avda. Angamos 0610, Casilla 1280 Antofagasta, Chile}


\begin{abstract}
We investigate thermodynamic schemes of charged BTZ-like black holes in arbitrary dimensions, namely higher-dimensional charged black holes in which the electromagnetic sector exhibits the same properties with that of the usual three-dimensional BTZ solution. We first present the Euclidean on-shell action in arbitrary dimensions, inserting a radial cutoff. We then extract the corresponding thermodynamic quantities from the semi-classical partition function in different ensembles and find that there exist two possible thermodynamic schemes, with different outcomes. Regarding the traditional scheme (scheme I), where the length cutoff is identified with the AdS radius, we show that charged BTZ-like black holes are super-entropic, namely they violate the reverse isoperimetric inequality conjecture, and their super-entropicity is strongly connected to a fundamental thermodynamic instability. This class of solutions is the first demonstration of super-entropic black holes which possess second-order critical points and, since thermodynamic instabilities always arise, it is not possible to physically interpret the corresponding van der Waals critical phenomenon in this scheme. In the second scheme (II) where the length cutoff is considered as an independent variable, namely the system respects the conjectured reverse isoperimetric inequality, we show that the solutions are thermodynamically stable in an ensemble where the length cutoff is 
kept fixed, and hence one can provide an explanation for the van der Waals critical phenomenon. Furthermore, in order to verify the consistency 
of the second scheme, we study the Joule-Thomson expansion and we extract the Joule-Thomson coefficient, 
the inversion temperature, the inversion curves, and the isenthalpic curves. The results indicate that this class of AdS black holes can be considered as interacting statistical systems. Additionally, in the $D \to 3$ limit we recover the conventional charge BTZ black holes, as well as their thermodynamic properties. Finally, we report a new thermodynamic instability for charged BTZ black holes, as well as their generalization to higher dimensions,   which implies that working in an ensemble with fixed length cutoff is mandatory to consistently examine the thermodynamic processes. 
\end{abstract}

%

 

\maketitle

\tableofcontents

\section{Introduction} \label{sec1:intro}

The investigation of black hole physics has opened   new gates of research that 
have expanded our understanding of strong gravity, quantum effects of 
gravitational fields, and black holes as thermal systems 
\cite{MTWbook1973,MukhanovWinitzki2007,BCH1973,Hawking1975Radiance,Addazi:2021xuf}. The 
study of   thermodynamic properties of asymptotically anti-de Sitter (AdS) 
black hole spacetimes is of prime 
importance, mainly for three reasons. Firstly, a well-defined 
description exists for the thermodynamics of black holes in AdS background,
unlike their cousins in asymptotically flat as well as asymptotically dS 
spacetimes \cite{HawkingPage1983, Ashtekar1984, Brown1994, Ashtekar2000}. 
Secondly, another motivation  comes from gauge/gravity duality (AdS/CFT 
correspondence \cite{Maldacena1998, 
PhysicsReports2000,Witten1998a,Witten1998b,GKP1998}), which establishes a 
connection between certain kinds of gauge field theories in $D$-dimensions and 
AdS gravity models in one higher dimensions. Thirdly, AdS black holes can 
imitate the thermodynamics of physical systems by assuming a negative dynamical 
cosmological constant ($\Lambda$) proportional to the thermodynamic pressure for 
them \cite{Kastor2009,Dolan2011CQG,CQG2017Review}.  \vspace{1mm}

There has been a resurgence of interest in the thermodynamics of AdS black 
holes in the last decade due to the extension of the thermodynamic phase space 
\cite{Kastor2009}, which has led to moving black hole physics forward and 
gaining a better understanding of their thermodynamics (for a nice review see 
\cite{CQG2017Review} and references therein). In this viewpoint, the 
cosmological constant is treated as a thermodynamic variable proportional to 
pressure and consequently the ADM mass $M$ is identified as the chemical 
enthalpy of the system  \cite{Kastor2009}, which leads to a  contribution of 
the pressure-volume term to the first law of thermodynamics. It is remarkable 
that without introducing the pressure-volume term in the thermodynamics of 
(A)dS black holes, the Smarr formula cannot be satisfied. In particular, the 
variation of $\Lambda$ in the first law of black hole thermodynamics naturally 
resolves the inconsistency between the Smarr formula and the traditional form of 
first law as \cite{Kastor2009,CQG2017Review}
\begin{gather} 
dM = TdS + VdP + ... \, \label{extended_first_law},\\ 
(D - 3)M = (D - 2)TS - 2PV + ... \, \label{extended_Smarr},
\end{gather}
where the neglecting part account for other possible work terms. In this 
context, the pressure and the thermodynamic volume in $D$-dimensions are given 
by \cite{Kastor2009}
\begin{equation} \label{PVterm}
P =  - \frac{\Lambda }{{8\pi {G_N}}} = \frac{{(D - 1)(D - 2)}}{{16\pi G_N 
{L^2}}}\,, \quad V = {\left( {\frac{{\partial M}}{{\partial P}}} 
\right)_{S,...}},
\end{equation}
where $G_N$ is the Newton constant and $L$ is the curvature radius of the
asymptotic AdS region. The above thermodynamic definition of volume can 
geometrically be realized by means of the Komar integral relation as 
\cite{Kastor2009}
\begin{equation} \label{geometric volume}
V =  {\int_{\partial {\Sigma _\infty }} {d{S_{ab}}\left( {{\omega ^{ab}} - 
\omega _{{\rm{AdS}}}^{ab}} \right) - \int_{\partial {\Sigma _{\rm{h}}}} 
{d{S_{ab}}{\omega ^{ab}}} } },
\end{equation}
where $d{S_{ab}}$ is the volume element normal to the co-dimension 2 surface 
$\partial \Sigma $, and the Killing potential ${\omega ^{ab}}$ is defined 
using the Killing vector (${\xi ^a}$) as ${\xi ^a} = {\nabla _b}{\omega 
^{ab}}$. This quantity is interpreted as an effective volume inside the event 
horizon, since it provides a measure of the volume excluded from the spacetime 
by the black hole horizon, and moreover it is useful to access the volume of 
black holes in asymptotically flat backgrounds by considering the limit 
$\Lambda \to 0$ \cite{Kastor2009,CQG2017Review,RevIso2011}. In general, the 
extended phase space thermodynamics suggests that every dimensionful 
parameter in the theory has a thermodynamic interpretation and thus it 
should be treated as thermodynamic variable \cite{Kastor2010}. This property 
circumvents the inconsistency between the first law and the corresponding Smarr 
relation for any modified theory of general relativity \cite{CQG2017Review,Kastor2010}.  \vspace{1mm}

Extended phase space thermodynamics, that is rooted in the identifications 
(\ref{PVterm}), leads to a number of interesting results 
\cite{Kastor2009,Dolan2011CQG,CQG2017Review,RevIso2011,
Kastor2010,Dolan2011, 
Dolan2014CQG,KubiznakMann2012,Mann2012Altamirano,Mann2014TriplePoint, 
HennigarMann2017PRL,Astefanesei2019,DHM2020PRD,
Astefanesei2022, 
MultiCritical2022aJHEP,MultiCritical2022b,MultiCritical2022c, 
WeiLiu2015PRL,RevIso2013dS,IsoCP2014,DS2022,HE2014,
HE2017,HE2018a,
HE2018b,HE2019,HE2021,JT2017,JT2018a,JT2018b,JT2019,Kastor2019,
McInnes2021} (see 
ref. \cite{CQG2017Review} for a nice review). A remarkable property that emerges 
from this modern treatment of black hole thermodynamics is the discovery of a 
number of thermodynamic phase transitions in certain AdS black holes 
\cite{CQG2017Review,KubiznakMann2012,Mann2012Altamirano,Mann2014TriplePoint, 
HennigarMann2017PRL,Astefanesei2019,DHM2020PRD,
Astefanesei2022}, some of them close to realistic phase transitions in nature. 
This field of research still offers further new surprises; for example, 
$n$-tuple critical points associated with multicritical phase transition have 
been disclosed very recently 
\cite{MultiCritical2022aJHEP,MultiCritical2022b,MultiCritical2022c}. Additionally, as another remarkable result,  the 
extended phase-space thermodynamics using the thermodynamic definition of 
volume (\ref{PVterm}) has provided a better understanding of isoperimetric 
inequalities, which lead to generalizing this topic to (A)dS black hole 
spacetimes \cite{RevIso2011,RevIso2013dS}. It could also shed new light on the 
statistical nature of black holes by introducing the concept of number density 
of black hole molecules to gain insight into their microscopic structure 
\cite{WeiLiu2015PRL}. This context is also useful to carefully define a suitable 
scheme for the thermodynamics of black holes. A  nice example towards this 
direction can be found in   \cite{Frassino2015,Appels2020,JT2021}, where the 
authors offer a new view for the correct thermodynamic scheme of BTZ black 
holes when an $U(1)$ charge parameter is included.  \vspace{1mm}

Motivated by the aforementioned intriguing findings, in this work  we 
argue that we can obtain extra information  about various thermodynamic 
schemes of black holes via the extended phase space thermodynamics. We would 
like to focus on a gravitational theory which entails a hard cutoff in the black 
hole geometry even in AdS background. Black holes having a length cutoff in 
their geometry are thermodynamically challenging to address, especially when 
they are studied in the extended phase space thermodynamics, since such property 
leads to the emergence of different thermodynamic schemes. Since such a (length) 
cutoff is a dimensionful parameter, it essentially has a thermodynamic 
interpretation in the view of the extended phase space context. It is 
interesting, as well as instructive, to follow what the extended phase space 
thermodynamics implies about such theories. \vspace{1mm}

In fact, there is a good and 
important example of such theories in AdS background, the Einstein-Maxwell 
gravity in (2+1)-dimensions, which admits the so-called charged (rotating) BTZ 
black holes \cite{MTZ2000,Chan1995Mann} and extensively studied in a number of 
papers \cite{BTZstudy2006,Setare2008,BTZstudy2009,BTZstudy2010,BTZstudy2011, 
BTZstudy2022,Gonzalez:2011dr}. In order to find a finite ADM mass for charged BTZ black holes 
one needs to insert a radial cutoff ($R_0$) to regularize the divergent 
behavior, which basically corresponds to enclosing the system in a finite region 
(a circle with radius $R_0$) \cite{MTZ2000,Setare2008} and subsequently leads to 
the question of how we should treat this parameter in thermodynamic 
considerations. After the birth of the extended phase space thermodynamics, some 
developments took place which led to improving our understanding of the 
thermodynamics of charged BTZ black hole solutions 
\cite{Frassino2015,Appels2020,JT2021,BTZmasive2018,Ghosh2020,Xu2020,Johnson2020, 
Mo2017}.  In this regard, studies on charged BTZ black holes with(out) a 
rotation parameter have revealed that there exist two possible thermodynamical 
schemes and the traditional one, that we call scheme I, is probably not the 
correct one \cite{Frassino2015,Appels2020,JT2021}. Three important conclusions 
can be drawn from these studies:
\begin{itemize}

    \item In the first scheme (I), where the enclosing radius $R_0$ is 
identified with the AdS radius $L$, charged BTZ black holes are super-entropic 
\cite{Frassino2015} (namely they violate the conjectured reverse isoperimetric 
inequality \cite{RevIso2011}) and thermodynamically unstable since the specific 
heat at constant volume is always negative \cite{Johnson2020}. In the second 
scheme (II), where the radial cutoff is treated as an independent variable, the 
reverse isoperimetric inequality is saturated \cite{Frassino2015} (hence black 
holes are sub-entropic) and no fundamental thermodynamic instability arises 
\cite{Johnson2020}. 
	
	\item Using the concept of number density of black hole molecules to 
measure the microscopic degrees of freedom of the system \cite{WeiLiu2015PRL}, 
it was shown that charged (rotating) BTZ black holes are associated with 
repulsive interactions among the microstructures \cite{Ghosh2020,Xu2020}, 
indicating they are interacting statistical systems in agreement with the 
previous results out of the context of extended phase space thermodynamics 
\cite{BTZstudy2006,BTZstudy2009,BTZstudy2011}. 
	
	\item The existence of Joule-Thomson expansion in any thermodynamic system 
indicates that the underlying microstructures are interacting. However, charged 
BTZ black holes in the first scheme (with $R_0=L$) do not exhibit the 
thermodynamic process of Joule-Thomson expansion and behave like an ideal gas, 
in strong disagreement with the other results 
\cite{BTZstudy2006,BTZstudy2009,BTZstudy2011,Ghosh2020,Xu2020}. Within scheme 
II, the Joule-Thomson expansion arises which is a direct result of the fact 
that these systems are interacting \cite{JT2021} in line with previous results 
\cite{BTZstudy2006,BTZstudy2009,BTZstudy2011,Ghosh2020,Xu2020}. A heat engine 
study of charged BTZ black holes also confirmed that the second scheme enjoys 
internal consistency unlike the first scheme \cite{Mo2017}.	
\end{itemize}

In this work we are interested in studying higher-dimensional black hole 
solutions that exhibit some of the BTZ features, which are called ``BTZ-like''
black holes in the literature \cite{Hendi2011}. In particular, in this solution 
subclass  the electromagnetic sector   is adjusted in order for  the gauge 
potential ($A_\mu$) in all spacetime dimensions to acquire the same logarithmic 
form with that of   static solutions of Maxwell electrodynamics in 
three-dimensions, hence the electric field of a point charge in any dimension 
will have a form proportional to $1/r$ \cite{Gonzalez2009}. Such a property makes the black hole 
metric function similar to that of a usual three-dimensional  charged BTZ 
black hole, which in turn leads to the generalization of many of the 
thermodynamic properties of a usual charged BTZ black hole to higher 
dimensions. 
We mention however that  only the properties related to 
the electromagnetic sector of a charged BTZ black hole are generalized to 
higher dimensions, and that is why one adds the word ``-like'', namely to 
denote the partial character of the  generalization.\vspace{1mm}

The theory that we are interested in consists of Einstein gravity minimally 
coupled to a particular nonlinear electrodynamics, whose $D \rightarrow 3$ 
limit coincides with the three-dimensional Einstein-Maxwell gravity (see Sec. 
\ref{sect2:action} for the action principle and static solutions of this 
model). A number of interesting features of charged BTZ black holes that are 
absent in higher dimensions originate from the special behavior of Maxwell's 
electrodynamics in two spatial dimensions, including super-entropicity 
\cite{Frassino2015}, the emergence of logarithmic divergence and the necessity 
of enclosing system for renormalization of the solution \cite{MTZ2000} etc. 
\cite{Setare2008,BTZstudy2006,BTZstudy2009,BTZstudy2010,BTZstudy2011,
BTZstudy2022,Johnson2020,Xu2020,Mo2017,Ghosh2020}. One can reproduce the 
effects of electromagnetic sector of charged BTZ black holes in higher 
dimensions, as suggested in refs. \cite{Hendi2011,Gonzalez2009}, by implementing a special 
class of the so-called power Maxwell invariant (PMI) theory of electrodynamics 
as the matter field \cite{HassaineMartinez2007,HassaineMartinez2008} whose 
Lagrangian is given by
\begin{equation} \label{PMI Lagrangian}
{\mathcal{L}_{{\text{PMI}}}} = {( - \mathcal{F})^s}, \quad (\mathcal{F} = 
{F_{\mu \nu }}{F^{\mu \nu }} \, \text{with} \,
{F_{\mu \nu }} = {\partial _{[\mu }}{A_{\nu ]}}).
\end{equation}
In  \cite{Hendi2011,Gonzalez2009}  it was shown that Einstein gravity has to be 
minimally coupled with the PMI Lagrangian (\ref{PMI Lagrangian}) upon setting 
$s=(D-1)/2$ in order for building an approximate analogue of charged BTZ black 
holes living in higher dimensions.\footnote{This type of minimal coupling may also be useful for higher-dimensional, charged generalizations of non-singular $3D$ vortices presented in refs. \cite{Edery2021,Edery2022a,Edery2022b}.} It is expected that the same features to 
happen in higher dimensions, however at the same time, due to the additional 
degrees of freedom in higher dimensions, we also expect richer and more 
complicated physics compared to the conventional 3D charged BTZ black holes, as 
will be shown in this work. It should be emphasized that power Maxwell invariant 
theory was originally invented to extend the conformal invariance to higher 
dimensions if the power $s$ in the PMI Lagrangian is set
as $s=D/4$ \cite{HassaineMartinez2007,HassaineMartinez2008}, in a way 
reminiscent of the Maxwell action which enjoys conformal invariance in 
four-dimensions. It is easy to show that the corresponding stress-energy tensor 
is traceless in any dimension. Consequently, it is possible to have a 
conformally invariant Maxwell source for gravitating (black hole) systems in 
higher dimensions. Adopting this idea, coupling gravity to PMI electrodynamics 
as a matter source beyond the special choice of $s=D/4$ may resulted in some 
interesting consequences (for example, see 
\cite{Hendi2011,DH2021,Mazharimousavi2022,HendiVahidinia2013} and references therein). In this 
work, we concentrate on the special case of $s=(D-1)/2$, in which the results 
are drastically different and cannot be obtained by taking the $s \to (D-1)/2$ 
limit of the Einstein-PMI gravity theory with a general $s$.  \vspace{1mm}

In this manuscript we are interested in   investigating the charged BTZ-like 
black holes \cite{Hendi2011,Gonzalez2009} as the higher-dimensional generalization of the 
conventional charged BTZ black holes and bring some new perspectives on the 
thermodynamics of them, which can be generalized to the other types of black 
holes having a length cutoff in their geometry (e.g., by coupling gravity to 
nonlinear electrodynamics theories as a matter source in three-dimensions). We 
further develop the theory suggested in ref. \cite{Hendi2011} and generalize the 
planar \textit{charged BTZ-like black hole} solutions to spherically and 
hyperbolically symmetric geometries in flat as well as (A)dS 
backgrounds\footnote{Note that, in flat and dS backgrounds, only spherically 
symmetric black hole solutions are allowed.} and extensively examine the 
solutions in the extended phase space via two possible thermodynamic schemes. 
Using the evaluation of Euclidean on-shell action, we show that a length cutoff 
is required in the geometry of charged BTZ-like black holes. As will be shown in 
this research, the resulting charged BTZ-like black hole solutions in AdS are 
actually super-entropic in a particular scheme (I) where $R_0=L$, like their 
three-dimensional cousins (charged BTZ black holes \cite{Chan1995Mann,MTZ2000}), 
but the thermodynamic instabilities are challenging to address here. To our 
knowledge, this class of solutions is the first demonstration of super-entropic 
black holes which possesses second-order critical point and it is theoretically 
interesting to investigate how they exhibit $P-v$ criticality. While, in the 
other scheme (II) where $R_0 \ne L$, a totally different class of thermodynamic 
black holes is obtained which behaves physically if the length cutoff is not 
allowed to fluctuate (via a fixed $R_0$ ensemble). This provides further 
insights about the conventional charged BTZ black holes and a new thermodynamic 
instability associated with the varying length cutoff is reported. Although, we 
mainly focus on the AdS black hole thermodynamics, we also discuss a 
thermodynamic limit to asymptotically flat black hole spacetimes with the help 
of the extended phase space context. This study trigger this idea that we can 
learn more about black holes as thermodynamic systems via the extended phase 
space context when a cutoff is included in their geometry.\footnote{Analyzing black holes in the extended phase space can potentially lead to a deeper understanding of these objects. For example, see the recent work of ref. \cite{ElSayed2023} about extended black hole thermodynamics in five-dimensional minimal gauged supergravity.} It should be noted 
that the $D \to 3$ limit of all the outcomes of this paper recovers those of 
conventional charged BTZ black hole studied in two different schemes in refs. 
\cite{Frassino2015,Johnson2020,JT2021}.\vspace{1mm}  

After this introduction, in Sec. \ref{sect2:action}, we study the action 
principle of the model under consideration which consists of Einstein gravity 
minimally coupled to Maxwell Lagrangian with the power $s = (D-1)/2$ and the 
static solutions are critically discussed. In Sec. 
\ref{sect3:renormalization}, we present the Euclidean on-shell action of the 
black hole solutions in detail in both the fixed charge and the fixed potential 
ensembles and then the corresponding conserved charges are reviewed. After 
that, in Sec. \ref{sect4:thermodynamics}, the charged BTZ-like black holes as 
thermodynamic systems are studied in two possible schemes. Next, in Sec. 
\ref{sect5:RII}, the connection between the conjectured reverse isoperimetric 
inequality and thermodynamic instabilities are extensively studied as a new 
example for examining the instability conjecture of refs. 
\cite{Johnson2020,Cong2019Mann}. In sections \ref{sect6:criticality} and 
\ref{sect7:JT}, we concentrate on the critical phenomena and the Joule-Thomson 
expansion of the black hole solutions in order to seek internal consistency in 
a particular scheme. Finally, we summarize the results and finish our paper 
with some conclusions in Sec. \ref{sect8:conclusion}.

\section{Action principle and charged BTZ-like black holes in arbitrary 
dimensions} \label{sect2:action}

The Einstein's theory of general relativity can minimally be coupled with the 
special case of PMI electrodynamics (\ref{PMI Lagrangian}) with $s=(D-1)/2$ to 
build the so-called charged BTZ-like black holes in higher dimensions, as 
discussed in the previous section. Hence, for our purpose, the bulk action is 
given by \cite{Hendi2011,Gonzalez2009}
\begin{equation} \label{bulk action}
I_\text{Bulk} =  - \frac{1}{2\kappa}\int_\mathcal{M} {{d^D}x\sqrt { - g} \left[ 
{R - 2\Lambda  + {{( - \mathcal{F})}^{(D - 1)/2}}} \right]},
\end{equation}
where $\kappa = {8 \pi {G_N}}$ and hereafter we set $G_N=c=\hbar=1$. Note that, for the PMI Lagrangian (\ref{PMI Lagrangian}) which ours in (\ref{bulk action}) is a subclass of it, one can put the minus sign inside the parentheses, ${( - \mathcal{F})^s}$, or outside the parentheses, $- {( \mathcal{F})^s} $, which both cases have been considered by several authors separately \cite{HassaineMartinez2007,HassaineMartinez2008,Hendi2011,DH2021,Gonzalez2009,Mazharimousavi2022,HendiVahidinia2013,Zhang2023}. Coupling gravity to $- {( \mathcal{F})^s} $ is well defined when the power of Maxwell invariant, $s$, only takes positive integers but not fractional numbers (because $\cal F$ is not a positive definite operator). For the case of $s=(D-1)/2$, this means the theory is well defined only for odd dimensions. While, for the case with ${(- \mathcal{F})^s} $, fractional powers of $(-\cal F)$ are allowed provided that either no magnetic field is present ($\bf{B}=0$) \cite{Hendi2011} or the condition $\textbf{E}^2 > \textbf{B}^2$ is satisfied (since ${\cal F} = {F_{\mu \nu }}{F^{\mu \nu }} =  2({\bf{B}^2}-{{\bf{E}}^2})$, so ${(- \mathcal{F})^s}$ is well defined for the mentioned conditions on the magnetic field). This leads to a well-defined coupling of gravity with this special subclass of nonlinear electrodynamics not only in odd dimensions but also in even dimensions, which results in static black hole solutions in arbitrary dimensions with an electromagnetic logarithmic term in their metric function.\footnote{Note that, since we are dealing with static charged black holes without any magnetic fields, only the electrostatic field contributes in the Faraday tensor ${F_{\mu \nu }}$. Assuming the usual static homogeneous ansatz for the gauge field, ${A_\mu } =\phi(r)\delta _\mu ^0$, one finds ${\cal F} = {F_{\mu \nu }}{F^{\mu \nu }} =  -
2 A_0^2$, proving that the electromagnetic part of the Lagrangian in (\ref{bulk action}) as a special case of ${( - \mathcal{F})^s}$ is always positive definite everywhere \cite{DH2021}. We also highlight refs. \cite{Mazharimousavi2020CQG,Amirabi2021}, in which the authors have considered an alternative form of the theory as $ - \alpha {\left| {{F_{\mu \nu }}{F^{\mu \nu }}} \right|^s}$ that is well defined in any dimension. Regarding our model in (\ref{bulk action}), one can simply uses $ {\left|- \mathcal{F} \right| ^{(D - 1)/2}}$ instead of ${{( - \mathcal{F})}^{(D - 1)/2}}$ to circumvent the issue of fractional powers in even dimensions. Assuming this (coupling gravity to ${\left|- \mathcal{F} \right| ^{(D - 1)/2}}$), in static situations, the results remain the same as ours in the present paper.} Note that only one of these conditions is required for even dimensions, but the same physics is observed in either even or odd dimensions generally, as will be demonstrated in this paper. However, one can simply uses $ {\left|- \mathcal{F} \right| ^{(D - 1)/2}}$ instead of ${{( - \mathcal{F})}^{(D - 1)/2}}$ to circumvent the issue of fractional powers in even dimensions without any extra assumptions. Furthermore, assuming ${\mathcal{L}_{{\text{PMI}}}} = {( - \mathcal{F})^s}$ for the Lagrangian of PMI theory, the properties of causality and unitarity \cite{Shabad2011} are satisfied provided that $s \ge 1$ \cite{DH2021}, which is trivially valid for our case, i.e. $s=(D-1)/2$, if $D \ge 3$. Instead, assuming the case of $- {( \mathcal{F})^s} $, one can show that the theory satisfies causality and unitarity conditions only for (integer) odd values of $s$, which cannot be satisfied if $s=(D-1)/2$. All these guarantee that a well-defined action principle in any dimension exists for the model under consideration (\ref{bulk action}). \vspace{1mm}

Varying the bulk action (\ref{bulk action}) with respect to the dynamical metric 
$g_{\mu \nu}$ and the gauge potential $A_\mu$ leads to
\begin{eqnarray} \label{variation}
\delta {I_\text{Bulk}} &=&  - \frac{1}{{16\pi}}\int_\mathcal{M} {{d^D}x\sqrt { 
- g} \big[ {{G_{\mu \nu }} + \Lambda {g_{\mu \nu }} - {T_{\mu \nu }}} 
\big]\delta {g^{\mu \nu }}} \nonumber \\
&&+ \frac{1}{{8\pi}}\int_{\partial \mathcal{M}} {{d^{D - 1}}x\sqrt { - \gamma } 
} {n^\alpha }{\gamma ^{\mu \nu }}\delta {g_{\mu \nu ,\alpha }} \nonumber \\
&&- \frac{{D - 1}}{{8\pi }}\int_\mathcal{M} {{d^D}x\sqrt { - g} } {\nabla _\mu 
}\big[{\mathcal{F}^{(D - 3)/2}}{F^{\mu \nu }}\big]\delta {A_\nu } \nonumber \\
&& + \frac{{D - 1}}{{8\pi}}\int_{\partial \mathcal{M}} {{d^{D - 1}}x\sqrt { - 
\gamma } } {( - \mathcal{F})^{(D - 3)/2}}{n_\mu }{F^{\mu \nu }}\delta {A_\nu },
\end{eqnarray}
where $\gamma_{\mu \nu }$ is the induced metric of the boundary, $n_\mu$ is a 
radial unit vector pointing outwards, and $G_{\mu \nu }$ and $T_{\mu \nu }$ are 
the Einstein and the stress-energy tensors, respectively. The first and third 
lines of the r.h.s. of eq. (\ref{variation}) represent the field equations of 
motion as
\begin{equation} \label{field eqs_GR}
{G_{\mu \nu }} + \Lambda {g_{\mu \nu }} - \frac{1}{2}{g_{\mu \nu }}{( - 
\mathcal{F})^{(D - 1)/2}} - (D - 1){F_{\mu \lambda }}{F_\nu }^{\,\lambda }{( - 
\mathcal{F})^{(D - 3)/2}} = 0,
\end{equation}
\begin{equation} \label{field eqs_EM}
{\nabla _\mu }\big[{\mathcal{F}^{(D - 3)/2}}{F^{\mu \nu }}\big] = 0.
\end{equation}
From eq. (\ref{variation}), it is obvious that in order to have a well-posed 
action principle, the bulk action (\ref{bulk action}) have to be supplemented 
by the following surface terms
\begin{equation} \label{GHY term}
{{I}_{{\text{GHY}}}} =  - \frac{1}{{8\pi }}\int_{\partial \mathcal{M}} {{d^{D - 
1}}x\sqrt { - \gamma } K},
\end{equation}
and
\begin{equation} \label{EM surface term}
I_{{\text{surface}}}^{{\text{(EM)}}} =  - \frac{{D - 1}}{{8\pi}}\int_{\partial 
\mathcal{M}} {{d^{D - 1}}x\sqrt { - \gamma} } {( - \mathcal{F})^{(D - 
3)/2}}{n_\mu }{F^{\mu \nu }}{A_\nu },
\end{equation}
where ${ I}_\text{GHY}$ is the so-called Gibbons-Hawking-York boundary (GHY) 
term \cite{GHY1972,GHY1977} with $K = \frac{1}{{\sqrt \gamma  }}{n^\mu 
}{\partial _\mu }\sqrt \gamma  $ as the trace of extrinsic curvature of 
boundary ($\partial {\cal M}$) and $I_{{\text{surface}}}^{{\text{(EM)}}}$ is 
the surface term needed for canceling the boundary term arisen from the 
electromagnetic sector (the last surface integral in eq. (\ref{variation})). 
Therefore, the total action for evaluating the finite on-shell action in the 
next section may be written as
\begin{equation} \label{total action}
{I} = {{I}_\text{Bulk}} + {{I}_\text{GHY}} 
+I_{{\text{surface}}}^{{\text{(EM)}}},
\end{equation}
which is suitable for studying the system in the fixed charge (canonical) 
ensemble. The electromagnetic surface term in eq. (\ref{variation}) can be also 
eliminated by imposing a boundary condition as ${\left. {\delta {A_\nu }} 
\right|_{\partial \mathcal{M}}} = 0$, which defines a fixed potential ensemble 
(also known as the grand canonical ensemble). We present the on-shell action 
for both ensembles in Sec. \ref{sect3:renormalization}.  \vspace{1mm}

In order to solve the field equations, we make use of static ansatz for both 
the electrodynamics and the gravity sectors. Assuming the static homogeneous 
ansatz of ${A_\mu } =\phi(r)\delta _\mu ^0$ for the gauge field, the 
electromagnetic field equations (\ref{field eqs_EM}) reduces to a simple 
differential equation as $r\partial _r^2\phi (r) + {\partial _r}\phi (r) = 0$ 
in any dimension \cite{Hendi2011}, which admits the following solution
\begin{equation} \label{em_sol}
\phi (r) = {c_1}\ln r + {c_2} = q\ln r - q\ln {R_0} +c_3= q\ln \left( 
{\frac{r}{{{R_0}}}} \right) +c_3.
\end{equation}
As seen, we have set $c_1=q$ and $c_2=-q \ln R_0 +c_3$ for the integration 
constants (the reason for splitting $c_2$ into two parts will be clarified in a 
moment). Although, the second integration constant ($c_2$) does not affect the 
resultant electric field ($E=\nabla \phi(r)=q/r$) as well as the strength field 
tensor $F_{\mu \nu}$ (the only nonzero components are given by 
$F_{tr}=-F_{rt}=q/r$), but its logarithmic part ($-q \ln R_0$) is necessary to 
preserve the gauge invariance property (because the physical quantity is the 
difference of the gauge potential). In fact, the logarithmic term  $-\ln R_0$ 
leads to asymptotic region can be defined by enclosing the system within a 
large hypersurface of radius $R_0$. Consequently, the field will not diverge 
asymptotically, leading to a gauge invariant definition for the thermodynamic 
potential. The radial parameter $R_0$ also appears as a cutoff in 
renormalization procedure (as it will be shown in Sec. 
\ref{sect3:renormalization}) and finds an interesting interpretation in the 
context of the extended phase space thermodynamics (which is the subject of 
Sec. \ref{sect4:thermodynamics}). On the other hand, in order to the norm of 
the gauge field ($A_\mu A^\mu$) remains always finite at the horizon (denoted 
by $r=r_+$), we need to impose the regularity at the horizon by setting 
$A_t(r_+)= 0$. The regularity condition is also essential for evaluating 
semi-classical partition function via Euclidean formalism \cite{Horowitz2011} 
(see Sec. \ref{sect3:renormalization}). This is the reason why the constant 
$c_3$ was introduced in eq. (\ref{em_sol}) and it is simply obtained by 
imposing $A_t(r_+)= 0$. Thus, the gauge vector field is given by
\begin{equation} \label{gauge vector}
{A_\mu } = q\left[ {\ln \left( {\frac{r}{{{R_0}}}} \right) - \ln \left( 
{\frac{{{r_ + }}}{{{R_0}}}} \right)} \right]\delta _\mu ^0.
\end{equation}
Note that $R_0$ cannot be subtracted and consequently eliminated because the 
system is enclosed in a hypersphere of radius $R_0$ but, in principle, one can 
take the $R_0 \to \infty$ limit while the ratio $r/R_0=1$ is kept fixed at the 
boundary. As a result, the gauge-invariant electric potential is found as
\begin{equation} \label{potential}
\Phi  = {\left. {{A_\mu }{\chi ^\mu }} \right|_{r \to R_0 }} - {\left. {{A_\mu 
}{\chi ^\mu }} \right|_{r \to {r_ + }}} = -q\ln \left( {\frac{{{r_ + 
}}}{{{R_0}}}} \right),
\end{equation}
which certainly is gauge invariant since the asymptotic region is now defined 
by $r \to R_0$. As will be explored in this work, setting $R_0=L$ (i.e, 
identifying the radial cutoff with the radius of AdS space) that is common in 
literature leads to some pathological results while setting $R_0 \ne L$ leads 
to whole different thermodynamic black holes. \vspace{1mm}

For the dynamical metric, we now consider the static ansatz having the 
maximally symmetric horizons with sectional
curvature $k = 0, \pm 1$ as
\begin{equation} \label{metric}
d{s^2} =  - f(r)d{t^2} + \frac{{d{r^2}}}{{f(r)}} + {r^2}\bigg( {dx_1^2 + 
\frac{{{{\sin }^2}(\sqrt k {x_1})}}{k}\sum\limits_{i = 2}^{D - 2} 
{dx_i^2\prod\limits_{j = 2}^{i - 1} {{{\sin }^2}{x_j}} } } \bigg).
\end{equation}
The line element of the horizon represents a closed hypersurface with positive 
($k = +1$), zero ($k = 0$), or negative ($k = -1$) constant curvature of 
$k(D-1)(D-2)$ and volume $\Sigma_{(k)}$.\footnote{For spherical black holes in 
$D$-dimensional spacetimes, $\Sigma_{(k=+1)}$ is given by ${\Sigma _{(k =  + 
1)}} = \frac{{2{\pi ^{(D - 1)/2}}}}{{\Gamma \left[ {(D - 1)/2} \right]}}$.} 
Putting the above ansatz into the gravitational field equations (\ref{field 
eqs_GR}) results in the following metric function
\begin{equation} \label{metric function_nocf}
f_b(r) = k + \frac{{{r^2}}}{{{L^2}}} - \frac{m}{{{r^{D - 3}}}} - \frac{{{2^{(D 
- 1)/2}}{q^{D - 1}}}}{{{r^{D - 3}}}}\ln\left(r \right),
\end{equation}
and, for the time being, we have not imposed a radial cutoff for enclosing the 
system in a large hypersurface of radius $R_0$ since the metric function 
approaches that of AdS space as $r \to \infty$ if $D \ge 4$. However, sooner or 
later one finds that, like the case of conventional charged BTZ black holes 
\cite{MTZ2000,Setare2008}, adding a radial cutoff ($R_0$) is mandatory for 
different reasons, as listed below.
\begin{itemize}
	\item Firstly, it is not physical that the electromagnetic sector is 
confined in a hypersurface of radius $R_0$, but the gravitational sector 
extends to infinity. 
	
	\item  Secondly, the computation of Euclidean on-shell action and 
subsequently the semi-classical partition function reveals that without 
assuming an upper bound (denoted by $R_0$) on the radial component, it is not 
possible to get a finite result and the theory exhibits divergent terms in any 
case. We address this issue in Sec. \ref{sect3:renormalization} in detail. 
	
	\item Thirdly, we confirmed that if one naively ignored the radial cutoff 
in the metric function like eq. (\ref{metric function_nocf}), the electric 
potential would never be obtained in the gauge-invariant form of eq. 
(\ref{potential}) through the standard thermodynamic relations and 
thermodynamic quantities will also not satisfy the Smarr formula. This issue 
will be discussed in Sec. \ref{sect3:renormalization}.
	
\end{itemize}

For the aforementioned reasons, following \cite{MTZ2000}, the length parameter 
$R_0$ is introduced as the second integration constant for handling the 
divergence and is naturally identified with the one already introduced in the 
gauge potential (\ref{gauge vector}), i.e.,\footnote{In order to convert the 
convention of this paper to that of MTZ's paper \cite{MTZ2000}, it is enough to 
replace $q$ (in this paper)  with $Q/2$ to obtain the charged BTZ black hole 
solution of \cite{MTZ2000} in three-dimensions.}
\begin{equation} \label{metric function}
f(r) = k + \frac{{{r^2}}}{{{L^2}}} - \frac{m}{{{r^{D - 3}}}} - \frac{{{2^{(D - 
1)/2}}{q^{D - 1}}}}{{{r^{D - 3}}}}\ln \left( {\frac{r}{{{R_0}}}} \right).
\end{equation}
Again, a natural interpretation of the parameter $R_0$ is that the 
higher-dimensional charged BTZ-like black hole systems are enclosed in a 
hypersurface of radius $R_0$, leading to a finite ADM mass. The first clue to 
the interpretation of $R_0$ is inferred from the metric function (\ref{metric 
function}). In particular, it is obvious that a limit to asymptotically flat 
background of this class of solutions (via $\Lambda \to 0$ or equivalently $L 
\to \infty$) is not possible whenever $R_0=L$. This suggests that $R_0$ should 
be treated as an independent parameter to access the asymptotically flat limit. \vspace{1mm}

 In addition, the $D \to 3$ limit of the metric function (\ref{metric 
function}) recovers that of the well-known charged BTZ black hole spacetime 
\cite{MTZ2000} if we set $k=0$ (the case that studied in ref. 
\cite{Hendi2011}), but in this case ($k=0$) the coordinates $x_i$'s are not 
periodic that is an important difference compared with the 3D charge BTZ black 
hole.\footnote{For the case of planar black holes ($k=0$), the geometry of the 
horizon is Ricci flat, as it is obvious from the context.} Hence, the charged 
BTZ-like black holes with spherical geometry for the horizon are natural 
higher-dimensional generalizations of the conventional charged BTZ black holes. 
As usual, the largest root of the metric function, $f(r_+)=0$, with positive 
slope specifies the location of the event horizon, $r_+$. One can also confirms 
that the Kretschmann scalar diverges only at the origin ($r=0$), indicating an 
existence of essential singularity.  \vspace{1.2mm}

\textbf{Charged BTZ-like Black holes in de Sitter (dS) space.} This class of black hole spacetimes also exist (only for the case of spherical symmetry, $k=1$) as an exact solution to (\ref{bulk action}) with positive cosmological constant and can be obtained through the analytic continuation $L \to i L$ of the metric function (\ref{metric function}).

\section{Euclidean action and conserved charges} \label{sect3:renormalization}

In this section we present the Euclidean on-shell action (${\bar I}_E$) and 
subsequently the semi-classical partition function ($Z$) for the static AdS 
black holes solutions of the theory (\ref{bulk action}), in both 
the fixed charge and the fixed potential ensembles. We then generalize these results to the charged BTZ-like black holes in dS space. We show that introducing a 
hard cutoff, here $R_0$, on the radial integration is mandatory to get a finite 
result. This length cutoff will find a natural interpretation through the 
extended phase space thermodynamics in Sec. \ref{sect4:thermodynamics}. The 
implications of assuming that the black hole system is not enclosed in a 
hypersurface of radius $R_0$ (using the metric function (\ref{metric 
function_nocf})) will also be discussed. We should emphasize that the $D \to 3$ 
limit of the computation of Euclidean on-shell action provides strong proof 
indicating that a renormalization length scale is mandatory in the geometry of 
the charged BTZ black holes.\vspace{1mm}

In Euclidean formulation, the semi-classical black hole partition function can 
be evaluated by a path integral in Euclidean signature (through the Wick 
rotation the generating functional) over spacetime metric ($g_{\mu \nu}$) and 
matter fields (here $A_\mu$) as \cite{GHY1977}
\begin{equation} \label{path integral}
Z = \int {\mathcal{D}g\mathcal{D}A{e^{ - {I_E}[g,A]}}}  \approx {e^{ - {{\bar 
I}_E}}},
\end{equation}
where ${I}_E=-i I$ is the total action in the Euclidean signature and ${\bar 
I}_E$ is the Euclidean on-shell action. The Euclidean metric, which is obtained 
by performing a temporal Wick rotation ($t \to -i t_E$), is given by
\begin{equation}
ds_E^2 = f(r)dt_E^2 + f{(r)^{ - 1}}d{r^2} + {r^2}d\Sigma _{(k)}^2,
\end{equation}
where $d\Sigma _{(k)}^2$ represents spherical ($k=+1$), planar ($k=0$), and 
hyperbolic ($k=-1$) horizon geometries of the spacetime metric (\ref{metric}) 
with volume $\Sigma _{(k)}$. In this formalism, the imaginary time $t_E$ is 
naturally associated with the inverse temperature $\beta$ since regularity 
condition near the horizon requires the Euclidean time to be periodic, i.e. 
${t_E} \sim {t_E} + \beta $, which results in
\begin{equation} \label{temperature}
{\beta ^{ - 1}} = T={\left. {\frac{{f'(r)}}{{4\pi }}} \right|_{r = {r_ + }}} = 
\frac{{(D - 3)k}}{{4\pi {r_ + }}} + \frac{{(D - 1){r_ + }}}{{4\pi {L^2}}} - 
\frac{{{2^{(D - 1)/2}}{q^{D - 1}}}}{{4\pi r_ + ^{D - 2}}},
\end{equation}
in agreement with the definition of surface gravity for obtaining Hawking 
temperature \cite{Hawking1975Radiance}. As usual in relativistic field 
theories, the dominant contribution to the path integral is obtained by 
substituting the classical solutions of the action (the so-called saddle-point 
approximation). In the path integral (\ref{path integral}), the gauge field and 
the dynamical metric must asymptotically flow to zero and the pure 
AdS\footnote{By ``pure AdS'', we mean the standard AdS space without a black 
hole.}, respectively, in order to have a finite result. This justifies why a 
radial cutoff, $R_0$, is necessary for the gauge field and the metric in eqs. 
(\ref{gauge vector}) and (\ref{metric function}) in arbitrary dimensions 
including $D=3$.  \vspace{1mm}

 In order to evaluate the Euclidean on-shell action, we make use of the 
subtraction method \cite{HawkingPage1983}, in which the on-shell action of the 
AdS background without black hole and matter fields (referred to as 
$I_E^{{\text{(AdS)}}}$) is subtracted from that of the black hole spacetime 
under consideration (referred to as $I_E^{{\text{(BH)}}}$). In Euclidean 
formulation, for the bulk action and the GHY action, one generally obtains 
\begin{equation} \label{Euclidean bulk}
{I_{{\text{Bulk}}}} =  - \frac{1}{{16\pi }}\int_0^\beta  {d{t_E}} \int_{{\Sigma 
_{(k)}}} {{d^{D - 2}}\vec x} \int^{{R_0}} {dr\sqrt g {\mathcal{L}_E}[g,A]},
\end{equation}
and
\begin{eqnarray} \label{Euclidean GHY}
{I_{{\text{GHY}}}} &=&  - \frac{1}{{8\pi {G_d}}}\int_0^\beta  {d{t_E}} 
\int_{{\Sigma _{(k)}}} {{d^{D - 2}}\vec x\,{{\left[ {{n^\mu }{\partial _\mu 
}\sqrt \gamma  } \right]}_{r = {R_0}}}}  \nonumber \\
&=&  - \frac{{{\Sigma _{(k)}}\beta }}{{16\pi }}{\left[ {{r^{D - 2}}{\partial 
_r}f(r) + 2(D - 2){r^{D - 3}}f(r)} \right]}_{r = {R_0}},
\end{eqnarray}
where $\gamma$ is the trace of the induced metric of the boundary in Euclidean 
signature. The lower radial bound in the Euclidean bulk action for the black 
hole must be $r_+$ while for the pure AdS space it is set equal to zero. The 
pure AdS space as the thermal background is obtained by setting $m=q=0$ in the 
metric function (\ref{metric}), referred to as $f_0(r)$, and its period is 
denoted by $\beta_0$. In order to have the same geometry at $r=R_0$ for both 
the black hole and the AdS space, we demand ${\left. {\beta \sqrt {f(r)} } 
\right|_{r = {R_0}}} = {\left. {{\beta _0}\sqrt {{f_0}(r)} } \right|_{r = 
{R_0}}}$ which fixes the temperature at the boundary 
\cite{HawkingPage1983,Witten1998b}, yielding
\begin{equation} \label{beta condition}
{\beta _0} = \beta \left[ {1 - \frac{{m{L^2}}}{{2{r^{D - 1}}}} + {\mathcal{O}} 
\big({r^{ - 2(D - 1)}}\big)} \right]_{r=R_0}.
\end{equation}
Finally, using eqs. (\ref{Euclidean bulk}) and (\ref{Euclidean GHY}), 
subtracting the Euclidean on-shell bulk action of the pure AdS from that of the 
black hole leads to
\begin{equation} \label{bulk difference}
I_{{\text{Bulk}}}^{{\text{(BH)}}} - I_{{\text{Bulk}}}^{{\text{(AdS)}}} = 
\frac{{{\Sigma _{(k)}}\beta }}{{16\pi }}\left[ {kr_ + ^{D - 3} - \frac{{r_ + 
^{D - 1}}}{{{L^2}}} + {2^{(D - 1)/2}}{q^{D - 1}}\ln \left( {\frac{{{r_ + 
}}}{{{R_0}}}} \right)} \right].
\end{equation}
As we can see, we cannot take the $R_0 \to \infty$ limit at the end, 
nevertheless the parameter $R_0$ will act as a thermodynamic variable in Sec. 
\ref{sect4:thermodynamics}. Turning our attention to ${I_{{\text{GHY}}}}$, as 
usual, one expects that the GHY action in AdS backgrounds has a zero 
contribution to the total on-shell action 
\cite{HawkingPage1983,Witten1998b,Charmousis2011}. Interestingly, unlike the 
case of black hole spacetimes in AdS space, here the GHY action has a nonzero 
contribution and a nontrivial term survives due to the logarithmic part in the 
electromagnetic sector.\footnote{Even in massive gravity theory \cite{Cai:2012db} in which the 
metric function includes a term proportional to $r$ and diverges at asymptotic 
region faster than the logarithmic term in our case (\ref{metric function}), 
the contribution of GHY boundary action in the subtraction method is zero 
\cite{CQG2020}. The reason is that the divergence in the case under 
consideration stems from the matter field but not the gravitational theory.} In 
fact, the difference between the Euclidean GHY boundary action of the black 
hole and that of pure AdS along with fixing the temperature at infinity 
(\ref{beta condition}) yields
\begin{equation} \label{GHY difference}
I_{{\text{GHY}}}^{{\text{(BH)}}} - I_{{\text{GHY}}}^{{\text{(AdS)}}} = 
\frac{{{\Sigma _{(k)}}\beta }}{{16\pi }}{2^{(D - 1)/2}}{q^{D - 1}}.
\end{equation} 
Here, we should emphasize that if one utilizes a metric function without 
enclosing the system, as eq. (\ref{metric function_nocf}), then it turns out 
that
the same results as eqs. (\ref{bulk difference}) and (\ref{GHY difference}) are 
still obtained (with some differences in details) in which $R_0$ here will be a 
mandatory cutoff for regulating the divergence that emerged in the radial 
integration due to the behavior of the vector field.\footnote{Regarding this 
case, fixing the temperature at the boundary now yields
$${\beta _0} = \beta \left[ {1 - \frac{{m{L^2}}}{{2{R_0^{D - 1}}}} 
+\frac{2^{(D-1)/2}q^{D-1}L^2}{2} \frac{\ln R_0}{R_0^{D-1}}  +{\mathcal{O}}  
\left({R_0^{ - 2(D - 1)}}\right)} \right].$$
} In conclusion, there is no way to escape the divergent contributions unless 
the system is contained within a bounded region. Using these ingredients, we 
can now present the total on-shell action of the charged BTZ-like black holes 
in the fixed charge and the fixed potential ensembles, as summarized below. \vspace{1mm}

\textbf{Fixed charge (canonical) ensemble.} Fixing the electric charge at 
infinity is simply obtained by adding the surface term (\ref{EM surface term}) 
to the bulk action which leads to a well-posed action principle. Therefore, the 
Euclidean action of the black hole spacetime consists of $I_E^{{\text{(BH)}}} = 
I_{{\text{Bulk}}}^{{\text{(BH)}}} + I_{{\text{GHY}}}^{{\text{(BH)}}} + 
I_{{\text{surface}}}^{{\text{(EM)}}}$ while for that of the AdS space (without 
black hole) is   $I_E^{{\text{(AdS)}}} = I_{{\text{Bulk}}}^{{\text{(AdS)}}} + 
I_{{\text{GHY}}}^{{\text{(AdS)}}}$. The electromagnetic surface term (\ref{EM 
surface term}) of the total black hole action, 
$I_{{\text{surface}}}^{{\text{(EM)}}}$, is computed as \footnote{If one uses a 
vector field without an enclosing parameter $R_0$, it turns out that a 
mandatory radial cutoff will appear again and the same result as eq. 
(\ref{EM_on-shell}) is obtained.}
\begin{equation} \label{EM_on-shell}
I_{{\text{surface}}}^{{\text{(EM)}}} = \frac{{{\Sigma _{(k)}}\beta }}{{16\pi 
}}(D - 1){2^{(D - 1)/2}}{q^{D - 1}}\ln \left( {\frac{{{r_ + }}}{{{R_0}}}} 
\right).
\end{equation}
Now, using eqs. (\ref{bulk difference}) and (\ref{GHY difference}) along with 
the surface term (\ref{EM_on-shell}), the canonical on-shell action is computed 
as
\begin{eqnarray} \label{on-shell action_can}
\bar I_E^{{\text{(can)}}} &=& I_E^{{\text{(BH)}}} - I_E^{{\text{(AdS)}}} 
\nonumber \\
&=& \frac{{{\Sigma _{(k)}}\beta }}{{16\pi }}\left\{ {kr_ + ^{D - 3} - \frac{{r_ 
+ ^{D - 1}}}{{{L^2}}} + {2^{(D-1)/2}{q}^{D - 1}}\left[ {1 - (D - 2)\ln \left( 
{\frac{{{r_ + }}}{{{R_0}}}} \right)} \right]} \right\},
\end{eqnarray}
in which $T$, $P$, and $Q$ are the constants of the ensemble. The corresponding 
(Helmholtz) free energy is simply obtained as $F=-T \ln Z=T \bar 
I_E^{\text{(can)}}$. This result agrees with the thermodynamic definition of 
$F=M-TS$ and can be proved by directly calculating the thermodynamic quantities 
$M$, $T$, and $S$ using different approaches.  \vspace{1mm}

\textbf{Fixed potential (grand canonical) ensemble.} Another way to get rid of 
the electromagnetic surface integral in the variation (\ref{variation}) is 
fixing the potential at infinity by imposing a boundary condition as ${\left. 
{\delta {A_\nu }} \right|_{\partial \mathcal{M}}} = 0$. So, there is no 
necessity for adding the surface term (\ref{EM surface term}), 
$I_{{\text{surface}}}^{{\text{(EM)}}}$, anymore. This choice defines the fixed 
potential ensemble in which the electric charge is allowed to fluctuate. In 
conclusion, the Euclidean on-shell actions of the black hole spacetime and the 
pure AdS consist of $I_E^{{\text{(BH)}}} = I_{{\text{Bulk}}}^{{\text{(BH)}}} + 
I_{{\text{GHY}}}^{{\text{(BH)}}} $ and  $I_E^{{\text{(AdS)}}} = 
I_{{\text{Bulk}}}^{{\text{(AdS)}}} + I_{{\text{GHY}}}^{{\text{(AdS)}}}$, 
respectively. The final result is obtained as
\begin{eqnarray} \label{on-shell action_grand}
\bar I_E^{{\text{(grand)}}} &=& I_E^{{\text{(BH)}}} - I_E^{{\text{(AdS)}}} 
\nonumber \\
&=& \frac{{{\Sigma _{(k)}}\beta }}{{16\pi }}\left\{ {kr_ + ^{D - 3} - \frac{{r_ 
+ ^{D - 1}}}{{{L^2}}} + {2^{(D - 1)/2}}{(-\Phi) ^{D - 1}}{{\left[ {\ln \left( 
{\frac{{{r_ + }}}{{{R_0}}}} \right)} \right]}^{1 - D}}\left[ {1 + \ln \left( 
{\frac{{{r_ + }}}{{{R_0}}}} \right)} \right]} \right\},
\end{eqnarray}
in which, in order to write everything in terms of the constants of fixed 
potential ensemble, the charge parameter $q$ has been replaced by the potential 
$\Phi$ via eq. (\ref{potential}). The corresponding free energy is simply 
obtained as $F=-T \ln Z=T \bar I_E ^{\text{(grand)}}$. This result agrees with 
the thermodynamic definition of $F=M-TS-\Phi Q$ by directly computing $M$, $T$, 
$S$, $\Phi$, and $Q$. \vspace{1mm}

\textbf{Conserved charges.} Having the Euclidean on-shell action of the black 
hole solution, one can straightforwardly extract the corresponding conserved 
quantities. The ADM mass can be obtained from both the fixed charge and the 
fixed potential partition functions. Assuming the fixed charge ensemble 
(\ref{on-shell action_can}), the ADM mass of the black hole is computed as 
\begin{eqnarray} \label{ADM mass}
M &=&  - {\partial _\beta }\ln Z = {\partial _\beta }{{\bar I}_E} \nonumber\\
&=& \frac{{{\Sigma _{(k)}}(D - 2)}}{{16\pi }}\bigg[ kr_ + ^{D - 3} + \frac{{r_ 
+ ^{D - 1}}}{{{L^2}}} - {{2^{(D - 1)/2}}{q^{D - 1}}\ln \left( 
{\frac{r_+}{{{R_0}}}} \right)} \bigg].
\end{eqnarray}
Now, as usual, the entropy is calculated as
\begin{equation}
S = \beta M - {\bar I_E} = \frac{{{\Sigma _{(k)}}}}{4}r_ + ^{D - 2},
\end{equation}
in agreement with the area law, $S=A/4$. The total charge of the spacetime can 
be obtained using the usual method of calculating the flux of the electrostatic 
field at infinity\footnote{To do that, one can compute the charge passing 
through a $(D-1)$-dimensional spacelike hypersphere ($\partial \Sigma$) with 
the same geometry as the event horizon at spatial infinity 
\cite{Carroll2004Book}. Following \cite{Carroll2004Book}, by implementing the 
Stokes' theorem, the finite charge of the black hole in the presence of any 
nonlinear electrodynamics is given by the generalized Gauss' law as
$$	Q =  - \frac{1}{{4\pi }}\int_{\partial \Sigma } {{d^{D - 2}}x  \sqrt \gamma 
 {n_\mu }{\sigma _\nu }\left( {\frac{{\partial {{\cal 
L}_{{\rm{NED}}}}}}{{\partial {\cal F}}}{F^{\mu \nu }}} \right)},$$
where the boundary ($\partial \Sigma$) has the metric $\gamma_{ij}$ with 
outward-pointing normal vector ${\sigma ^\mu } = \sqrt {f(r)} \delta _r^\mu$, 
and ${n^\mu } =  - \frac{1}{{\sqrt {f(r)} }}\delta _t^\mu$ is the unit normal 
vector associated with $\Sigma$. Finally, assuming ${{\cal L}_{{\rm{NED}}}}={{( 
- \mathcal{F})}^{(D - 1)/2}}$, the total charge of the black hole spacetime is 
calculated as eq. (\ref{charge}).} or by considering  the fixed potential 
ensemble (\ref{on-shell action_grand}) and then applying the thermodynamic 
relation $Q =  - {\left( {\frac{{\partial F}}{{\partial \Phi }}} \right)_\beta} 
= \frac{1}{\beta }\frac{{\partial {{\bar I}_E}}}{{\partial \Phi }}$ along with 
eq. (\ref{potential}), which results in
\begin{equation} \label{charge}
Q = \frac{{{\Sigma _{(k)}}(D - 1){2^{(D - 1)/2}}{q^{D - 2}}}}{{16\pi }}.
\end{equation}
One can also check that, using the partition function in the fixed charge 
ensemble, the electrostatic potential difference is found in agreement with eq. 
(\ref{potential}) as
\begin{equation}
\Phi  ={\left( {\frac{{\partial F}}{{\partial Q }}} \right)_\beta} = 
\frac{1}{\beta }\frac{{\partial {{\bar I}_E}}}{{\partial Q}} =  - q\ln \left( 
{\frac{{{r_ + }}}{{{R_0}}}} \right).
\end{equation}
It is straightforward to verify that the obtained thermodynamic quantities 
($M$, $T$, $S$, $\Phi$, and $Q$) satisfy the first law of black hole 
thermodynamics as $dM=TdS+\Phi dQ$ (written in the energy representation) but 
it is impossible to find a Smarr relation just by use of these thermodynamic 
variables. The extended phase space thermodynamics naturally resolves this 
issue by introducing new thermodynamic variables, as will be clarified in the 
next section.

\textbf{The Euclidean action approach in dS space.} Our findings can be extended to the charged BTZ-like black holes in dS space. To do so, first note that for black holes in dS space there is a cosmological horizon ($r_\text{cosmo}$) with a negative Hawking temperature which excludes thermodynamic equilibrium between the event and the cosmological horizons for any observer. For this reason, one needs to enclose the system in a thermal cavity with fixed temperature. Interestingly, the model under consideration already involves a radial cutoff ($R_0$) which acts as a cavity in dS space and that is why our previous calculations in AdS were more like calculations in dS space. Following the literature \cite{Simovic2019,Simovic2021}, we demand the temperature associated with the cavity wall (at $R_0$) remains fixed which results in thermodynamic equilibrium within the cavity. Then, by performing calculations using the subtraction method, it turns out that all the computations presented in the AdS space, are also valid for the dS space upon the analytic continuation $L \to i L$. Note, like the AdS case, the upper bound of the radial integration in the bulk and the surface integrals is $R_0$ for both the black hole and the pure dS (without the black hole) in the subtraction method. As a result, the corresponding thermodynamic quantities in dS are also obtained by the analytic continuation $L \to i L$ of the same quantities in AdS. \vspace{1mm}

In the next sections, we confine our attention to the extended phase space thermodynamics of charged BTZ-like black holes with AdS asymptote.

\section{Extended phase space thermodynamics and two different schemes} 
\label{sect4:thermodynamics}

Identifying the pressure as eq. (\ref{PVterm}) implies that the ADM mass $M$ 
(\ref{ADM mass}) has to be identified as the chemical enthalpy and the total 
on-shell action (\ref{on-shell action_can}) is associated with the Gibbs free 
energy, given by
\begin{equation} \label{Gibbs}
G =  - T\ln Z = \frac{{{\Sigma _{(k)}}}}{{16\pi }}\left\{ {kr_ + ^{D - 3} - 
\frac{{16\pi Pr_ + ^{D - 1}}}{{(D - 1)(D - 2)}} + {2^{(D - 1)/2}}{q^{D - 
1}}\left[ {1 - (D - 2)\ln \left( {\frac{{{r_ + }}}{{{R_0}}}} \right)} \right]} 
\right\}.
\end{equation}
The Gibbs free energy plays a key role in analyzing critical phenomena. It is 
not difficult to verify that the Gibbs free energy can be obtained by using the 
Legendre transform of the enthalpy, i.e., $G=M-TS$. Now through the standard 
relations in thermodynamics such as $S =  - {\left( {\frac{{\partial 
G}}{{\partial T}}} \right)_{P,Q}}$, $\Phi  = {\left( {\frac{{\partial 
G}}{{\partial Q}}} \right)_{T,P}}$, and $V = {\left( {\frac{{\partial 
G}}{{\partial P}}} \right)_{T,Q}}$, one can compute thermodynamic quantities in 
the Gibbs energy representation. However, we would rather work in the enthalpy 
representation (common in the literature) in which one can instead use $T = 
{\left( {\frac{{\partial M}}{{\partial S}}} \right)_{P,Q}}$, $\Phi  = {\left( 
{\frac{{\partial M}}{{\partial Q}}} \right)_{S,P}}$, and $V = {\left( 
{\frac{{\partial M}}{{\partial P}}} \right)_{S,Q}}$ to obtain thermodynamic 
quantities. The extended phase space thermodynamics, the same as conventional 
charged BTZ black hole spacetime, depends on what we interpret the parameter 
$R_0$ which results in two drastically different schemes. Here we investigate 
the extended black hole thermodynamics in both possible schemes.

\subsection{Scheme I ($R_0=L$)} \label{sect:sch1}

The first possible scheme, the usual one, is found by identifying 
$R_0=L$. For convenience, all the thermodynamic quantities are summarized here 
\begin{gather}
M = \frac{{(D - 2){\Sigma _{(k)}}}}{{16\pi }}\bigg[ k r_ + ^{D - 3}+ \frac{{r_ 
+ ^{D - 1}}}{{{L^2}}} -  {{2^{(D - 1)/2}}{q^{D - 1}}\ln \left( {\frac{r_+}{L}} 
\right)} \bigg], \quad Q = \frac{{{\Sigma _{(k)}}(D - 1){2^{(D - 3)/2}}{q^{D - 
2}}}}{{8\pi }}, \nonumber \\
\Phi  =  - q\ln \left( {\frac{{{r_ + }}}{L}} \right), \quad T = \frac{{(D - 
3)k}}{{4\pi {r_ + }}} + \frac{{(D - 1){r_ + }}}{{4\pi {L^2}}} - \frac{{{2^{(D + 
1)/2}}{q^{D - 1}}}}{{8\pi r_ + ^{D - 2}}}, \quad S = \frac{{{\Sigma 
_{(k)}}}}{4}r_ + ^{D - 2}.
\end{gather}
Regarding the extended phase space, for the pressure and the thermodynamic 
volume conjugate to pressure, we have
\begin{equation}
P = \frac{{(D - 1)(D - 2)}}{{16\pi {L^2}}},\,\,\,\,\,V = {\left( 
{\frac{{\partial M}}{{\partial P}}} \right)_{S,Q}} = \frac{{{\Sigma 
_{(k)}}}}{{D - 1}}\left( {r_ + ^{D - 1} - {2^{(D - 3)/2}}{L^2}{q^{D - 1}}} 
\right).
\end{equation}
It should be noted that a finite limit to asymptotically flat spacetime 
($\Lambda \to 0$ or $L \to \infty$) cannot be obtained for the ADM mass, the 
thermodynamic volume, and also the electrostatic potential difference. However, 
it is a matter of calculation to show that the first law and the corresponding 
Smarr relation take the following forms
\begin{equation}
dM = TdS + VdP + \Phi dQ,
\end{equation}
\begin{equation}
\frac{{D - 3}}{{D - 2}}M = TS - \frac{2}{{D - 2}}PV + \frac{{D - 3}}{{D - 
1}}\Phi Q,
\end{equation}
respectively. Note that the electromagnetic part of the Smarr formula is 
different form that of Reissner-Nordstr\"{o}m-AdS black holes since here the 
electric charge scales as $\left[ Q \right] = {[{\text{Length}]}^{(D - 2)(D - 
3)/(D - 1)}}$ (which follows from Eulerian scaling \cite{Kastor2009}). In the 
$D=3$ limit, one obtains a very simple form of $2PV=TS$ for the standard 
charged BTZ black holes which reminds us that of non-interacting ideal gas 
system. In the view of thermodynamics, an existence of Joule-Thomson expansion 
indicates that the system is interacting and this process is generally absent 
for non-interacting ones, as proved for the the case of charged BTZ black holes 
in this specific scheme in Ref. \cite{JT2021}. This signals a major 
inconsistency in this scheme because the other studies show that charged BTZ 
black holes are indeed interacting systems 
\cite{BTZstudy2006,BTZstudy2009,BTZstudy2011,Ghosh2020,Xu2020}. In the next 
section (\ref{sect5:RII1}) we also show that, in this scheme, the black hole 
system is always thermodynamically unstable, which means that identifying 
$R_0=L$ for the metric function (\ref{metric function}) leads to a pathological 
solution.

\subsection{Scheme II ($R_0 \ne L$)} \label{sect:sch2}

Following \cite{Frassino2015}, the second possible scheme can be established by 
identifying the length cutoff ($R_0$) as a new, independent thermodynamic 
parameter. The thermodynamic quantities in this scheme read
\begin{gather} 
M = \frac{{(D - 2){\Sigma _{(k)}}}}{{16\pi }}\bigg[ {k r_ + ^{D - 3} + 
\frac{{r_ + ^{D - 1}}}{{{L^2}}} - {2^{(D - 1)/2}}{q^{D - 1}}\ln \left( 
{\frac{r_+}{{{R_0}}}} \right)} \bigg], \quad Q = \frac{{{\Sigma _{(k)}}(D - 
1){2^{(D - 3)/2}}{q^{D - 2}}}}{{8\pi }}, \nonumber\\
\Phi  =  - q\ln \left( {\frac{{{r_ + }}}{{{R_0}}}} \right), \quad T = \frac{{(D 
- 3)k}}{{4\pi {r_ + }}} + \frac{{(D - 1){r_ + }}}{{4\pi {L^2}}} - \frac{{{2^{(D 
+ 1)/2}}{q^{D - 1}}}}{{8\pi r_ + ^{D - 2}}}, \quad S = \frac{{{\Sigma 
_{(k)}}}}{4}r_ + ^{D - 2}, \label{thermo_schemeII}
\end{gather}
implying that the thermodynamic volume and a new intensive quantity $K_0$ (the 
thermodynamic conjugate of $R_0$) are given by \footnote{A minor typo has 
occurred in the sign of the $D=3$ limit of eq. (\ref{K0}) in ref. 
\cite{Frassino2015} and it has been repeated many times in other papers. The 
importance of the sign of this quantity will be clarified in the next section 
while studying thermodynamic instabilities.}
\begin{gather}
P = \frac{{(D - 1)(D - 2)}}{{16\pi {L^2}}}, \quad V = {\left( {\frac{{\partial 
M}}{{\partial P}}} \right)_{S,Q,{R_0}}} = \frac{{{\Sigma _{(k)}}r_ + ^{D - 
1}}}{{D - 1}}, \label {PV_schemeII} \\
{K_0} = {\left( {\frac{{\partial M}}{{\partial {R_0}}}} \right)_{S,P,Q}} = 
\frac{{{\Sigma _{(k)}}(D - 2){2^{(D - 1)/2}}{q^{D - 1}}}}{{16\pi {R_0}}}. 
\label{K0}
\end{gather}
Unlike the previous scheme, a finite limit to asymptotically flat spacetime is 
simply obtained for all the thermodynamic quantities by taking $\Lambda \to 0$ 
($L \to \infty$). The above set of thermodynamic quantities satisfy the first 
law as
\begin{equation} \label{firstLawII}
dM = TdS + VdP + \Phi dQ + {K_0}d{R_0}.
\end{equation}
Now, the Smarr relation is found to be as
\begin{equation} \label{SmarrII}
\frac{{D - 3}}{{D - 2}}M = TS - \frac{2}{{D - 2}}PV + \frac{{D - 3}}{{D - 
1}}\Phi Q + \frac{1}{{D - 2}}{K_0}{R_0}.
\end{equation}
The variation of the parameter $R_0$ can be interpreted thermodynamically as: 
when the whole energy content of the black hole spacetime changes, the area 
(hypersurface of radius $R_0$) enclosing the system changes as well. In the 
$D=3$ limit, one obtains $2PV=TS+{K_0}{R_0}$ which indicates charged BTZ black 
holes are interacting statistical systems in complete agreement with the 
results so far found 
\cite{JT2021,BTZstudy2006,BTZstudy2009,BTZstudy2011,Ghosh2020,Xu2020}. We 
confirmed that no pathological behavior is observed in this scheme if one deals 
with an ensemble where $R_0$ is kept fixed, as will be further investigated in 
the next sections (\ref{sect5:RII2}, \ref{sect6:criticality}, and 
\ref{sect7:JT}).

\subsection{Alternative scheme? If $R_0$ is not a variable} 
\label{sect:sch3}

A question of interest is whether there exists another thermodynamic scheme 
besides the previous two. In fact, one can still treat the cosmological 
constant as a thermodynamic variable (which is the primary assumption of the 
extended phase space) while assuming the length cutoff $R_0$ to be fixed as an 
independent parameter of the theory (namely not a thermodynamic variable) to 
escape the relevant thermodynamic instabilities (see the next section 
\ref{sect5:RII} for more details). At the end, we will comment on why this 
scheme suffers from arbitrariness and is not acceptable within the context of 
the extended phase space thermodynamics. Assuming this scenario, there will be 
no more thermodynamic potential $K_0$ associated with $R_0$. The thermodynamic 
quantities remain the same as those presented in scheme II (Sec. 
\ref{sect:sch2}), i.e., eqs. (\ref{thermo_schemeII}) and (\ref{PV_schemeII}), 
although the two quantities $R_0$ and $K_0$ are no longer present. Now, the 
first law and the Smarr formula take the following forms
\begin{equation}
dM = TdS + VdP + \Phi dQ,
\end{equation}
\begin{equation}
\frac{{D - 3}}{{D - 2}}M = TS - \frac{2}{{D - 2}}PV + \frac{{D - 3}}{{D - 
1}}\Phi Q + \frac{{{\Sigma _{(k)}}}}{{16\pi }}{2^{(D - 1)/2}}{q^{D - 1}},
\end{equation}
 respectively. As will be clear in the next section \ref{sect5:RII}, there is 
no thermodynamic instability in this scheme (unlike that of associated with 
$R_0$ in scheme II) and well-defined descriptions are found for the critical 
phenomena and thermodynamic processes such as Joule-Thomson expansion exactly 
the same as the second scheme.  \vspace{1mm}
 
We should emphasize that this scheme has not to be taken rigorously. In fact, 
in order to preserve the standard definition of the thermodynamic volume as eq. 
(\ref{PV_schemeII}) in agreement with geometric definition (\ref{geometric 
volume}), one needs to introduce the new variable $R_0$, as shown for the case 
of three-dimensional charged BTZ black holes in ref. \cite{Frassino2015} in 
detail. For this reason, this scheme suffers arbitrariness and cannot be 
supported by the context of extended phase space thermodynamics.

\section{Reverse isoperimetric inequality and the instability conjecture} 
\label{sect5:RII}
The thermodynamic volume arisen from the extended black hole thermodynamics is 
conjectured to satisfy the so-called reverse isoperimetric inequality (RII) 
\cite{RevIso2011}. The RII conjecture for AdS black holes claims that 
\cite{RevIso2011}
\begin{equation} \label{RII}
{\cal R} \equiv {\left[ {\frac{{(D - 1)V}}{{{\Sigma _{(k)}}}}} 
\right]^{\frac{1}{{D - 1}}}}{\left( {\frac{{{\Sigma _{(k)}}}}{A}} 
\right)^{\frac{1}{{D - 2}}}} \ge 1,
\end{equation}
where $A$ is the area of the outer horizon and $V$ is the thermodynamic volume 
of the black hole. There are some counterexamples violating this conjecture, 
leading to what might be called super-entropic black holes 
\cite{UltraSpinningBH2015PRL,UltraSpinningBH2015JHEP,UltraSpinningBH2016NPB}. 
Violation of the RII conjecture means that the associated entropy exceeds the 
maximal bound implied by the inequality (\ref{RII}) or equivalently the 
extended black hole volume is no longer bounded from below 
\cite{UltraSpinningBH2015PRL}. Subsequent studies have indicated that violation 
of this conjecture in AdS black holes is somehow related to a new kind of 
fundamental thermodynamic instabilities \cite{Johnson2020,Cong2019Mann} and we 
show this is also valid for charged BTZ-like black holes.  

The isoperimetric ratio $\cal R$ could depend on the thermodynamic scheme one 
is dealing with. In what follows, we investigate the conjectured RII as well as 
the possibility of thermodynamic instabilities in both schemes. The local 
stability criteria can be examined in the energy representation as well as any 
possible Legandre transforms of the energy such as enthalpy, which is a bit 
different from the analyses have so far been done for the thermodynamic 
instabilities of super-entropic black holes. It turns out that the approach we 
present here is more powerful than studying the behavior of the specific heats, 
as will be clarified in this section.  \vspace{1mm}

At equilibrium, the energy $E$ is minimum, so $E$ must be a convex function of 
its natural extensive variables including the entropy $S$ and the volume $V$. 
Without loss of generality, we consider the system while its electric charge is 
held fixed. Hence, the local conditions of convexity for the energy become 
\cite{CallenBook}
\begin{equation} \label{stability_criteria_E}
{\left( {\frac{{{\partial ^2}E}}{{\partial {V^2}}}} \right)_S} \ge 0  \quad  
{\rm{and}}  \quad  {\left( {\frac{{{\partial ^2}E}}{{\partial {S^2}}}} 
\right)_V} \ge 0.
\end{equation}
The energy representation of black holes can be obtained using the Legendre transform of the enthalpy as $E=H-PV$, in which $H$ is actually the ADM mass $M$ in the extended phase space. In the extended phase space, the enthalpy appears naturally and it seems easier to work with enthalpy especially when $V$ and $S$ are not independent which often occurs in static black hole spacetimes. The entropy and the pressure are the natural extensive and intensive variables of the enthalpy, respectively, which imply that the local conditions of 
thermodynamic stability in the enthalpy representation are given by 
\cite{CallenBook}
\begin{equation} \label{stability_criteria_H}
{\left( {\frac{{{\partial ^2}H}}{{\partial {P^2}}}} \right)_S} \le 0  \quad  
{\rm{and}}  \quad  {\left({\frac{{{\partial ^2}H}}{{\partial {S^2}}}} 
\right)_P} \ge 0.
\end{equation}
These conditions are enough in order to investigate fundamental thermodynamic 
instabilities of the system.

\subsection{RII and thermodynamic instability in scheme I} \label{sect5:RII1}

Considering scheme I, using the relevant thermodynamic quantities presented in 
Sec. \ref{sect:sch1}, the isoperimetric ratio is found as
    \begin{equation} \label{IR_schI}
	\mathcal{R} = {\left[ {1 - \frac{{{2^{(D - 11)/2}}(D - 1)(D - 2){q^{D - 
1}}}}{{\pi Pr_ + ^{D - 1}}}} \right]^{\frac{1}{{D - 1}}}} < 1 \, .
    \end{equation}
As seen, under any circumstances, the RII conjecture is violated and black 
holes are super-entropic in this particular scheme. In the $D \to 3$ limit, the 
isoperimetric ratio (\ref{IR_schI}) reduces to that of the standard charged 
BTZ black holes, ${\mathcal{R}_{(D = 3)}} = \sqrt {1 - \frac{{{q^2}}}{{8\pi Pr_ 
+ ^2}}}  < 1$ \cite{Frassino2015}, as expected. In fact, the logarithmic term 
of the metric function (\ref{metric function}) is responsible for the 
super-entropicity when the length cutoff is identified with the AdS radius, 
$R_0=L$.  \vspace{1mm}

Now, we turn our attention to the connection between the violation of the RII 
conjecture and the emergence of fundamental thermodynamic instabilities in this 
scheme. For investigating thermodynamic instability of the system, we compute 
the stability criteria given in eq. (\ref{stability_criteria_H}) which result in
\begin{equation}
{\left( {\frac{{{\partial ^2}H}}{{\partial {P^2}}}} \right)_S}\, = 
\,\frac{{{\Sigma _{(k)}}{2^{(D - 11)/2}}(D - 2){q^{D - 1}}}}{{\pi {P^2}}},
\end{equation}
and
\begin{equation} \label{second}
{\left( {\frac{{{\partial ^2}H}}{{\partial {S^2}}}} \right)_P} = \frac{{{2^{(D 
- 1)/2}}{{(D - 2)}^2}{q^{D - 1}} + r_ + ^{D - 3}[16\pi Pr_ + ^2 - k(D - 2)(D - 
3)]}}{{{\Sigma _{(k)}}\pi {{(D - 2)}^2}r_ + ^{2D - 4}}}.
\end{equation}
As seen, the black hole system under consideration is thermodynamically 
unstable because the first stability condition in eq. 
(\ref{stability_criteria_H}), i.e. ${\left( {\frac{{{\partial ^2}H}}{{\partial 
{P^2}}}} \right)_S} \le 0,$ is always violated and takes positive values only. 
The same result is also obtained from the thermodynamic stability requirement 
of ${\left( {\frac{{{\partial ^2}E}}{{\partial {V^2}}}} \right)_S} \ge 0$ in 
the energy representation. The other second-order derivative, eq. 
(\ref{second}), which is related to the specific heat at constant pressure 
($C_P$), takes both positive and negative values depending on the choice of 
parameters. \vspace{1mm}

For the sake of completeness as well as later applications (in the next section 
\ref{sect6:criticality}), let us take a look at the specific heats, $C_P$ and 
$C_V$. The negativity of specific heats (here, at constant volume $C_V$ and at 
constant pressure $C_P$) as well as the other thermodynamic coefficients (such 
as isothermal compressibility $\kappa_T$ and isobaric expansivity $\beta_P$) 
signals local thermodynamic instability. The specific heat at constant pressure 
and the specific heat at constant volume are respectively given by
\begin{eqnarray} \label{c_p}
     {C_P} &=& T{\left( {\frac{{\partial S}}{{\partial T}}} \right)_P} 
\nonumber \\
     &=& \frac{{{\Sigma _{(k)}}\pi {{(D - 2)}^2}r_ + ^{2D - 4}T}}{{{2^{(D - 
1)/2}}{{(D - 2)}^2}{q^{D - 1}} - r_ + ^{D - 3}[k(D - 2)(D - 3) - 16\pi Pr_ + 
^2]}}
     \end{eqnarray}
and
\begin{eqnarray} \label{Cv}
{C_V} &=& T{\left( {\frac{{\partial S}}{{\partial T}}} \right)_V} \nonumber \\
&=& \frac{{{\Sigma _{(k)}}{2^{(D - 1)/2}}\pi {{(D - 2)}^3}r_ + ^{2D - 
4}T}}{{{2^{(D - 1)/2}}(D - 2)r_ + ^{D - 3}[16\pi r_ + ^2P - k(D - 2)(D - 3)] + 
{2^{D - 1}}{{(D - 2)}^3}{q^{D - 1}} - 512{q^{1 - D}}{{(\pi {r_ + }P)}^2}}}.
\end{eqnarray}
In the $D=3$ limit, which corresponds to the case of charged BTZ black holes 
(by setting $\Sigma _{(k)}=2 \pi$ and $k=0$), one finds $C_P > 0$ and $C_V < 0$ 
everywhere, in agreement with the results of \cite{Johnson2020}. In higher 
dimensions ($D \ge 4$), a richer behavior is observed, as expected from eqs. 
(\ref{c_p}) and (\ref{Cv}); for the most of parameter space we observe that 
$C_P > 0$ whenever $C_V < 0$. There are still some regions where 
both $C_P$ and $C_V$ are positive but we confirmed that the isothermal 
compressibility ($\kappa_T$) is always negative in such situations. An explicit 
example of such situation will be presented in the next section which happens 
during phase transitions. Therefore, the analyses of specific heats alone does 
not provide us information about the fundamental thermodynamic instability. \vspace{1mm}

In conclusion, the charged BTZ-like black holes are super-entropic and 
thermodynamically unstable in scheme I, which mean that identifying $R_0=L$ for 
the metric function (\ref{metric function}) leads to a pathological solution. 
The results confirms that violation of the RII conjecture is connected to the 
emergence of fundamental thermodynamic instabilities in support of the 
instability conjecture of refs. \cite{Johnson2020,Cong2019Mann}.

\subsection{RII and thermodynamic instability in scheme II}\label{sect5:RII2}

	Regarding the second thermodynamic scheme of the charged BTZ-like black 
holes presented in Sec. (\ref{sect:sch2}), one simply finds ${\cal R}  = 1$, 
showing that the RII conjecture is always respected (saturated in this case). \vspace{1mm}

	Now, we focus on the thermodynamic stability of the system in scheme II. 
Since the thermodynamic quantities $T$, $S$, and $P$ remain the same in both 
schemes, the specific heat at constant pressure $C_P$ is obtained the same as 
before, presented in eq. (\ref{c_p}), while the specific heat at constant 
volume $C_V$ vanishes since the thermodynamic volume $V$ and the entropy $S$ 
are not independent variables anymore in this scheme, i.e.,
\begin{equation}
	{C_V} = T{{\left({\frac{{\partial S}}{{\partial T}}} \right)}_{V}} = 0.
\end{equation}
On the other hand, the first local condition of thermodynamic stability in the 
enthalpy representation is satisfied trivially, i.e.,
    \begin{equation}
	{\left( {\frac{{{\partial ^2}H}}{{\partial {P^2}}}} \right)_S}\, = 0, 
	\end{equation}
while the second one remains the same as the one (\ref{second}) in scheme I. 
Obviously, the previous fundamental thermodynamic instability found in the 
first scheme does not arise here.  \vspace{1mm}

Furthermore, since the energy $E$ must be a convex function of its natural 
extensive variables including $R_0$, if the length cutoff of the system allows 
to fluctuate, which corresponds to dealing with an ensemble with fixed 
potential $K_0$, another condition for the local stability of the system 
besides those in eq. (\ref{stability_criteria_E}) is imposed as
\begin{equation} \label{convexity_R0}
\left( {\frac{{{\partial ^2}E}}{{\partial {R_0^2}}}} \right)_{S, V, Q} \ge 0.
\end{equation}
It is easy to show that the energy is not convex along this extensive variable 
because
\begin{equation} \label{instability_R0}
{\left( {\frac{{{\partial ^2}E}}{{\partial R_0^2}}} \right)_{S,V,Q}} =  - 
\frac{{{\Sigma _{(k)}}(D - 2){2^{(D - 1)/2}}{q^{D - 1}}}}{{16\pi R_0^2}} < 0 \,.
\end{equation}
In particular, for the conventional charged BTZ black holes in 
three-dimensions, one finds ${\left( {\frac{{{\partial ^2}E}}{{\partial 
R_0^2}}} \right)_{S,V,Q}} = - \frac{Q^2}{R_0^2}$. This is a new thermodynamic 
instability associated with the length cutoff even for the charged BTZ black 
holes (not reported before). To eliminate the thermodynamic instability 
associated with the length cutoff, there are two options. The first choice is to 
work in an ensemble where the length cutoff is kept fixed which leads to 
consistent results. The second choice is to assume the length cutoff is not a 
variable, so there will be no thermodynamic instability associated with this 
external parameter. If one assumes that the length cutoff is a fixed and 
independent parameter of the theory but not a variable, there will be no more 
the intensive/extensive pair of $K_0$ and $R_0$ in thermodynamic considerations. 
No thermodynamic instability associated with $R_0$ arises here since the 
convexity condition (\ref{convexity_R0}) is no longer relevant. The 
thermodynamic quantities remain the same as those in scheme II, i.e. eqs. 
(\ref{thermo_schemeII}) and (\ref{PV_schemeII}), and the isoperimetric ratio is 
found the same as before, i.e., ${\cal R}=1$. However, as mentioned before in 
Sec. \ref{sect:sch3}, this alternative scheme is not physically acceptable 
because the volume (\ref{PV_schemeII}) has obtained by the fact that the length 
cutoff $R_0$ is a thermodynamic variable.

\section{Critical phenomena} \label{sect6:criticality}

In this section we discuss the critical phenomena of charged BTZ-like black 
holes   in detail. Using the analyses of isobars and isotherms in 
the $T-r_+$ and the $P-r_+$ phase diagrams, it is shown that charged BTZ-like 
black holes in scheme I are the first demonstration of super-entropic black 
holes which possess second-order critical points, so it is theoretically 
interesting to investigate the associated critical behavior if exists any. We 
find that identifying $R_0=L$ in the metric function (\ref{metric function}) 
leads to a somewhat pathological class of thermodynamic black holes near the 
critical point. While, in the second scheme (II) where $R \ne L$, the same 
qualitative behavior for the critical phenomenon, like those previously found in 
many black hole spacetimes, is observed in a consistent way. \vspace{1mm}

The equation of state of the system is obtained from the Hawking temperature 
(\ref{temperature}), which takes the same form in both schemes if it is written 
down in terms of the horizon's radius, and is given by
\begin{equation} \label{pressure}
P = \frac{{(D - 2)T}}{{4{r_ + }}} - \frac{{(D - 2)(D - 3)k}}{{16\pi r_ + ^2}} + 
\frac{{{2^{(D - 1)/2}}(D - 2){q^{D - 1}}}}{{16\pi r_ + ^{D - 1}}}.
\end{equation}
In comparison with the equation of state of the van der Waals (vdW) fluid, the 
specific volume $v$ is identified with $v = \frac{{4{r_ + }}}{{D - 2}}$ (in the 
units where ${G_N}=c=\hbar=1$) to make contact with that of the vdW system 
having $P=\frac{T}{v}+ ...$ \cite{CQG2017Review}. No matter which variable we 
work with ($r_+$ or $v$), the results remain the same in any case. \vspace{1mm}

In order to study the criticality of a thermodynamic system, the first step is 
to find the corresponding critical point(s). The critical points occur at the 
inflection points of isotherms in the $P-v$ (or equivalently $P-r_+$) diagrams 
and can be found through the following relations
\begin{equation} \label{critical point criterion1}
{\left( {\frac{{\partial P}}{{\partial v}}} \right)_T} = {\left( 
{\frac{{{\partial ^2}P}}{{\partial {v^2}}}} \right)_T} = 0 \quad 
\leftrightarrow \quad {\left( {\frac{{\partial P}}{{\partial {r_ + }}}} 
\right)_T} = {\left( {\frac{{{\partial ^2}P}}{{\partial {r_ +^2 }}}} \right)_T} 
= 0 \,.
\end{equation}
Equivalently, the critical points can be determined by
\begin{eqnarray} \label{critical point criterion2}
{\left( {\frac{{\partial T}}{{\partial v}}} \right)_P} = {\left( 
{\frac{{{\partial ^2}T}}{{\partial {v^2}}}} \right)_P} = 0  \quad 
\leftrightarrow \quad {\left( {\frac{{\partial T}}{{\partial {r_ + }}}} 
\right)_P} = \left(\frac{{{\partial ^2}T}}{{\partial {r_ +^2}}} \right)_P= 0,
\end{eqnarray}
which means that the critical points are inflection points in the $T-v$ 
($T-r_+$) diagrams as well. These points occur at the spike-like divergence of 
the specific heat at constant pressure since ${\frac{{\partial T}}{{\partial 
{r_ + }}}}=0$. This can be manifested by
\begin{equation}
{C_P} =-T{\left( {\frac{{\partial ^2 G}}{{\partial T^2}}} \right)_P}= T{\left( 
{\frac{{\partial S}}{{\partial T}}} \right)_P}  = T\,\frac{{{{\left( 
{\frac{{\partial S}}{{\partial {r_ + }}}} \right)}_P}}}{{{{\left( 
{\frac{{\partial T}}{{\partial {r_ + }}}} \right)}_P}}}.
\end{equation}
So, using ${C_P} =  - T{\left( {{\partial ^2}G/\partial {T^2}} \right)_P}$ and 
Ehrenfest classification of phase transitions, one finds that every critical 
point, which satisfies the criterion (\ref{critical point criterion1}) or 
(\ref{critical point criterion2}), is of second-order type. \vspace{1mm}

It turns out that the equation of state (\ref{pressure}) possesses only one 
critical point and the corresponding critical radius is given by
\begin{equation}
{r_c} = {\left[ {\frac{{{2^{(D - 3)/2}}(D - 1)(D - 2){q^{D - 1}}}}{{(D - 3)k}}} 
\right]^{\frac{1}{{D - 3}}}}.
\end{equation}
Obviously, a critical point exists in $D \ge 4$ if and only if $k=+1$. The 
critical data including $r_c$ ($\propto v_c$), $T_c$, and $P_c$ remain the same 
in both schemes but the critical volume $V_c$ takes different values. It is 
verified that the critical phenomenon associated to this critical point is the 
well-known vdW-like behavior. In general, the standard vdW-like behavior in AdS 
black holes appears if equation of state admits one physical critical point 
($P_c$, $T_c$), for which a second-order, continuous phase transition occurs. 
The associated critical exponents at the critical point are the same as vdW 
fluid (i.e., $\alpha=0$, $\beta=1/2$, $\gamma=1$ and $\delta=3$). This behavior 
is simply confirmed by the characteristic swallowtail form of the corresponding 
$G-T$ diagrams for sub-critical isobars ($P<P_c$) or equivalently by the 
sub-critical isotherms ($T<T_c$) in the $G-P$ diagrams, for which a first-order 
phase transition takes place. Basically, one can extract the same information 
about the criticality by analyzing sub-critical isotherms ($T<T_c$) in $P-r_+$ 
($P-v$) diagrams or sub-critical isobars ($P<P_c$) in $T-r_+$ ($T-v$) diagrams. 
The resultant holographic interpretation is the small/large black hole phase 
transition \cite{KubiznakMann2012}. \vspace{1mm}

\begin{figure}
	\includegraphics[scale=0.55]{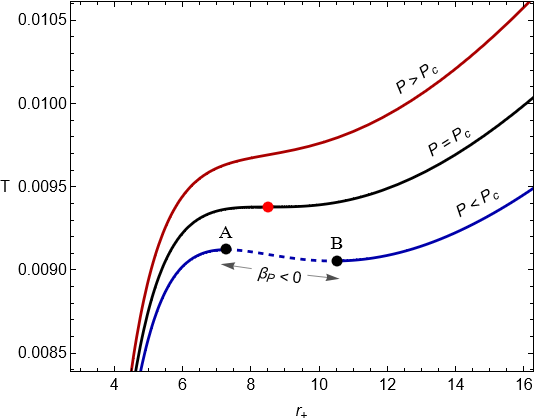}
	\hskip 1 cm
	\includegraphics[scale=0.55]{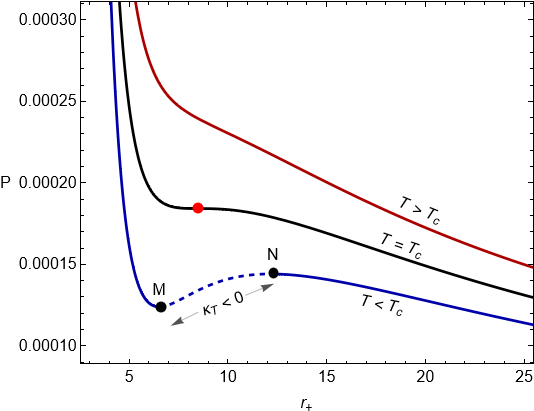}
	\caption{ \it{$T-r_+$ (left panel) and $P-r_+$ (right panel) phase 
diagrams for spherical charged BTZ-like black holes in $4$-dimensions,
for  fixed $q=1$ and $R_0=1000$.  The pressure of isobars 
in the $T-r_+$ diagram   and the temperature of isotherms in the 
$P-r_+$ diagram   decreases from top to bottom, while the red circles 
correspond to the critical point. The dashed curves correspond to unstable 
states having negative isobaric expansivity ($\beta_P < 0$) and negative 
isothermal compressibility ($\kappa_T < 0$), respectively.
	 The critical point corresponds to $r_c=8.4853$, $T_c=0.009378$, and 
$P_c=0.0001842$.}}
	\label{PV}
\end{figure}

Now, we can illustrate a typical example of the vdW-like phase transition in 
the model under consideration. In Fig. \ref{PV}, both the $T-{r_+}$ and 
$P-{r_+}$ phase diagrams are depicted for a typical four-dimensional charged 
BTZ-like black holes with spherical symmetry. These diagrams are valid for both 
schemes since the temperature and the pressure remain the same in any case, but 
the corresponding $G-T$ and $G-P$ diagrams depend on the thermodynamic scheme 
one is dealing with and will be discussed later. In Fig. \ref{PV}, the pressure 
of isobars in the $T-r_+$ diagram (left panel) and the temperature of isotherms 
in the $P-r_+$ diagram (right panel) decrease from top to bottom. The red 
circles correspond to critical point which is an inflection point in both the 
$T-r_+$ and the $P-r_+$ diagrams. This point exactly occurs at the spike-like 
divergence of the specific heat at constant pressure which is a sign for 
second-order phase transition and, according to Ehrenfest classification, it 
should be further analyzed by means of Gibbs free energy phase diagrams later. 
As seen in the right panel of Fig. \ref{PV}, the characteristic oscillatory 
feature of isotherms in the $P-r_+$ plane like those in vdW system is observed 
for isothermal curves having $T < T_c$. The part of the curve that has positive 
slopes (dashed line) corresponds to unstable states having negative isothermal 
compressibility, $\kappa_T < 0$, so it is unphysical and must be replaced by a 
line of constant pressure according to the Maxwell’s equal-area law. This 
signals the two-phase behavior associated with the occurrence of large/small 
black hole phase transition, which is further confirmed by the sub-critical 
isobars ($P<P_c$) in the $T-r_+$ diagram displayed in the left panel of Fig. 
\ref{PV}. Regarding the $T-r_+$ diagram, the part of the sub-critical isobaric 
curve ($P < P_c$) that has negative slope (dashed line) corresponds to unstable 
states having negative isobaric expansivity, $\beta_P < 0$. No phase transition 
is seen for high temperature isotherms ($T > T_c$) because of the absence of an 
oscillatory behavior. This is further confirmed by the high pressure isobars 
($P > P_c$). All these are quite analogous to the liquid/gas phase transition 
occurring in the vdW system. \vspace{1mm}

More detailed information about criticality can be extracted from the Gibbs 
free energy diagrams and analyzing $P-r_+$ and $T-r_+$ diagrams are not enough. 
For example, according to Ehrenfest classification, the order of the lowest 
differential of Gibbs free energy which shows a discontinuity at the critical 
pressure $P_c$ or equivalently at the critical temperature $T_c$ is the order 
of a phase transition. To be more specific, extra information about the isobars 
in the $T-r_+$ diagram and the isotherms in the $P-r_+$ diagram associated with 
Fig. \ref{PV} can be derived from the corresponding $G-T$ and $G-P$ diagrams, 
respectively. However, the Gibbs free energy diagrams depend on the scheme one 
is dealing with and, for this reason, in what follows we study them case by 
case. \\

\begin{figure}
	\includegraphics[scale=0.55]{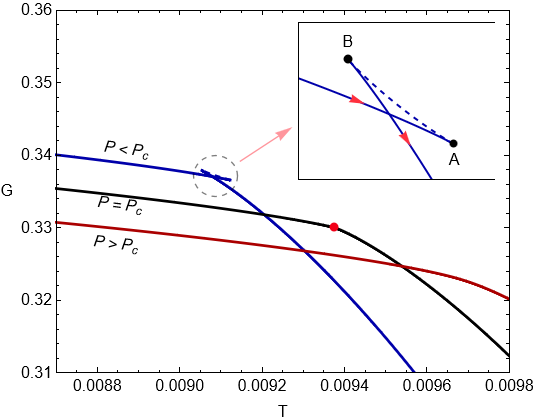}
	\hskip 1 cm
	\includegraphics[scale=0.56]{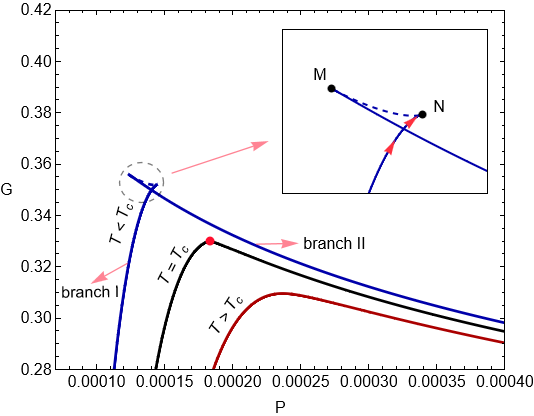}
	\caption{  \it{Scheme I: $G-T$ (left panel) and $G-P$ (right panel) 
phase diagrams for spherical charged BTZ-like black holes $4$-dimensions, for 
 fixed $q=1$ and $R_0=1000$.   The pressure of isobars in the $G-T$ 
diagram   and the temperature of isotherms in the $G-P$ diagram 
  increases from top to 
bottom, while the red circles 
correspond to the critical point. The red arrows in both diagrams indicate 
increasing horizon's radius $r_+$ (or equivalently increasing specific volume, 
$v$).  The dashed curves correspond to 
unstable states having negative isobaric expansivity ($\beta_P < 0$) and 
negative isothermal compressibility ($\kappa_T < 0$), respectively. However, the 
branch II of the sub-critical isotherm in the $G-P$ diagram also suffers from 
negative isothermal compressibility, which forbids any physical phase 
transition.
  The critical point corresponds to $r_c=8.4853$, $T_c=0.009378$, and 
$P_c=0.0001842$.}}
\label{Gibbs_sch1}
\end{figure}

\textbf{The analysis in scheme I.} Here we show that there is no way to explain 
the vdW-like critical phenomenon in scheme I because of the fundamental 
thermodynamic instability which shows itself in some way in the corresponding 
free energy phase diagrams. To show this, we concentrate on the Gibbs free 
energy diagrams, i.e., the $G-T$ and $G-P$ ones displayed in Fig. 
\ref{Gibbs_sch1}. As seen in the $G-T$ diagram (the left panel of Fig. 
\ref{Gibbs_sch1}), the swallowtail behavior (the triangle loop) is observed for 
the sub-critical isobar ($P < P_c$) which explicitly confirms a first-order 
phase transition since the preferred phase is determined by minimizing the 
Gibbs energy. The red arrows in the $G-T$ diagram, which show the preferred 
phase, indicate increasing both the temperature and the horizon's radius. So, 
it is inferred that a phase transition from a small black hole state to a large 
black hole state takes place. In this figure, the part of the triangle loop 
(dashed blue line) of the isobar with $P < P_c$, which is concave upward, 
corresponds to the dashed blue part of the corresponding isobar with $P < P_c$ 
in the $T-r_+$ diagram displayed in Fig. \ref{PV}. This part is unphysical 
since the corresponding isobaric heat capacity as well as isobaric expansivity 
are negative. In addition, at the critical point a second-order phase 
transition occurs since the conditions (\ref{critical point criterion2}) are 
satisfied which is further confirmed from ${C_P} =  - T{\left( {{\partial 
^2}G/\partial {T^2}} \right)_P}=\infty$ and, for isobars having $P > P_c$, a 
normal single-phase behavior is observed. So far everything seems fine but we 
noticed that any attempt to describe the vdW-like critical phenomenon in a 
consistent way leads to failure because the fundamental thermodynamic 
instability in this scheme (found in Sec. \ref{sect5:RII1}) makes one of the 
thermodynamic quantities $C_P$ (isobaric specific heat), $C_V$ (isochoric 
specific heat) or $\kappa_T$ (isothermal compressibility) always negative 
definite. This can be realized from the $G-P$ phase diagram in the right panel 
of Fig.\ref{Gibbs_sch1}, in which two parts of the triangle loop have positive 
concavity, meaning that the system cannot undergo a phase transition and 
selects thermodynamic coordinates that are in the direction of the red arrows 
(see the zoomed-in part of the right panel of Fig. \ref{Gibbs_sch1}). From this 
diagram, it is inferred that no phase transition takes place since the whole 
branch II is unstable and, although both the specific heats are positive, the 
isothermal compressibility is always negative definite which forbids any 
physical phase transition. All these prove that basically we are dealing with a 
pathological thermodynamic system.\\

\begin{figure}
	\includegraphics[scale=0.55]{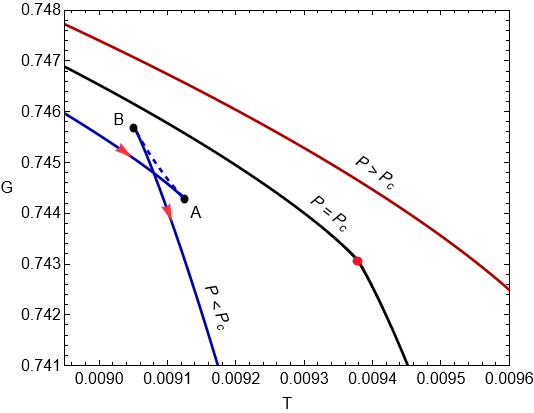}
	\hskip 1 cm
	\includegraphics[scale=0.53]{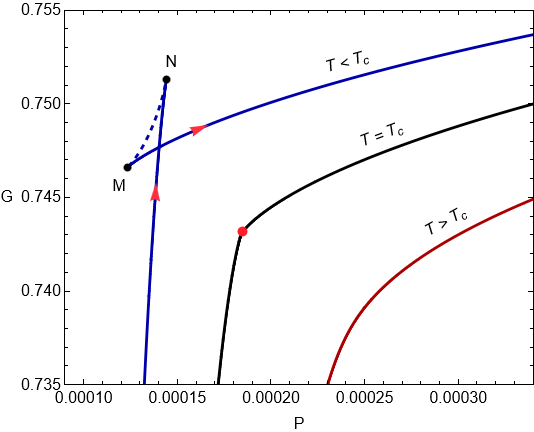}
	\caption{ \it{Scheme II: $G-T$ (left panel) and $G-P$ (right panel) 
phase diagrams for spherical charged BTZ-like black holes $4$-dimensions, for 
 fixed $q=1$ and $R_0=1000$.   The pressure of isobars in the $G-T$ 
diagram   and the temperature of isotherms in the $G-P$ diagram 
  increases from top to 
bottom, while the red circles 
correspond to the critical point. The red arrows in both diagrams indicate 
increasing horizon's radius $r_+$ (or equivalently increasing specific volume, 
$v$).  The dashed curves correspond to 
unstable states having negative isobaric expansivity ($\beta_P < 0$) and 
negative isothermal compressibility ($\kappa_T < 0$), respectively. However, 
the 
branch II of the sub-critical isotherm in the $G-P$ diagram also suffers from 
negative isothermal compressibility, which forbids any physical phase 
transition.
  The critical point corresponds to $r_c=8.4853$, $T_c=0.009378$, and 
$P_c=0.0001842$.}
 }
\label{Gibbs_sch2}
\end{figure}

\textbf{The analysis in scheme II.} In order to have a well-defined explanation 
for the vdW-like phase transition, the characteristic swallowtail behavior must 
be observed in both the $G-T$ and the $G-P$ diagrams. In both diagrams, only 
one part of the triangle loops must be concave upward which correspond to the 
unphysical (oscillatory) parts of the sub-critical isobars in the $T-r_+$ 
diagrams and the sub-critical isotherms in the $P-r_+$ diagrams displayed in 
Fig. \ref{PV}. In fact, in the second scheme, all these are satisfied exactly 
the same as vdW fluid. We display this in the $G-T$ diagram in the left panel 
of Fig. \ref{Gibbs_sch2}, which demonstrates the swallowtail behavior that is 
further confirmed by the $G-P$ diagram displayed in the right panel of this 
figure. If the pressure (temperature) is set in the range of $P < P_c$ 
($T<T_c$), by monotonically increasing the temperature (pressure) the system 
undergoes a first-order order phase transition. At the critical point a 
second-order phase transition takes place and for the isobars having $P > P_c$ 
or the isotherms having $T > T_c$ a single-phase behavior is observed. 
Obviously, the analyses of critical behavior from both the $G-T$ and the $G-P$ 
diagrams are in complete agreement with those of the $T-r_+$ and the $P-r_+$ 
diagrams, indicating a well-defined explanation for the vdW-like critical 
behavior. On the other hand, all the thermodynamic coefficients take physical 
values in this scheme which are a direct consequence of the absence of 
fundamental thermodynamic instabilities. Therefore, the analysis of critical 
behaviors in this scheme leads to consistent results.  

\section{Joule-Thomson expansion: A  consistency check  } \label{sect7:JT}

So far we disclosed that the study of thermodynamic processes is not physically acceptable in 
the first scheme due to the essential thermodynamic instabilities which arise 
from this pathological scheme by identifying $R_0=L$. In what follows, instead, we focus 
on the second scheme (II) and investigate the thermodynamics process of 
Joule-Thomson expansion in charged BTZ-like black holes for the sake of 
completeness, indicating that the results are physically consistent.  \vspace{1mm}

The Joule-Thomson expansion is an adiabatic irreversible process in which the 
temperature of a gas changes as it passes through a porous plug. The rate of 
the change of temperature with pressure during this isoenthalpic expansion ($dH 
\equiv dM=0$) is measured by the Joule-Thomson (JT) coefficient, $\mu=\left( 
\frac{\partial T}{\partial P} \right)_{M}$. A positive (negative) JT 
coefficient implies cooling (heating) of the system as the pressure falls. The 
curve separating regions with positive and negative JT coefficient is called 
the inversion curve on which $\mu=0$ and is plotted in the $T-P$ plane. The JT 
coefficient of the charged BTZ-like black holes can be calculated by 
differentiating the Smarr relation (\ref{SmarrII}) along with the help of the 
first law of thermodynamics (\ref{firstLawII}) for fixed values of $Q$ and 
$R_{0}$, given by
\begin{equation}\label{coeff}
\mu  = \frac{1}{{\left( {D - 2} \right)S}}\left[ {2P{{\left( {\frac{{\partial 
V}}{{\partial P}}} \right)}_M} - \frac{{\left( {D - 2} \right)\left( {D - 3} 
\right)}}{{D - 1}}Q{{\left( {\frac{{\partial \Phi }}{{\partial P}}} \right)}_M} 
+ DV} \right],
\end{equation}
where the partial derivatives in eq. (\ref{coeff}) are evaluated from the 
differentiation of the potential and volume in eqs. (\ref{thermo_schemeII}) and 
(\ref{PV_schemeII}) rewritten in terms of the entropy. After all, the JT 
coefficient for the charged BTZ-like black holes is given by
\begin{equation}
\mu  = \frac{1}{{D - 1}}\frac{{{\mu _N}}}{{{\mu _D}}},
\end{equation}
where
\begin{eqnarray}
{\mu _N} &=&  - {4^{\frac{{2D - 1}}{{D - 2}}}}\sqrt 2 D\left( {D - 3} 
\right)k\,q{\left( {\frac{S}{{{\Sigma _{(k)}}}}} \right)^{\frac{{D + 1}}{{D - 
2}}}} - {2^{\frac{{17D - 14}}{{2D - 4}}}}\pi qP{\left( {\frac{S}{{{\Sigma 
_{(k)}}}}} \right)^{\frac{{D + 3}}{{D - 2}}}}\nonumber\\
&& + {4^{\frac{{D + 1}}{{D - 2}}}}\left[ { - {{32}^{{\frac{{{D^2} - 2D + 
4}}{{2(D - 2)}}}}} + {2^{\frac{{{D^2}}}{{2\left( {D - 2} \right)}}}}D} 
\right]{q^D}{\left( {\frac{S}{{{\Sigma _{(k)}}}}} \right)^{\frac{4}{{D - 
2}}}},\nonumber\\
{\mu _D} &= & - {2^{\frac{{2D}}{{D - 2}}}}\sqrt 2 \left( {D - 2} \right)\left( 
{D - 3} \right)k\,q{\left( {\frac{S}{{{\Sigma _{(k)}}}}} \right)^{\frac{D}{{D - 
2}}}} - {2^{\frac{{13D - 10}}{{2D - 4}}}}\pi qP{\left( {\frac{S}{{{\Sigma 
_{(k)}}}}} \right)^{\frac{{D + 2}}{{D - 2}}}}\nonumber\\
&&+ {2^{{\frac{{{D^2} - 2D + 12}}{{2(D - 2)}}}}}\left( {D - 2} 
\right){q^D}{\left( {\frac{S}{{{\Sigma _{(k)}}}}} \right)^{\frac{3}{{D - 2}}}}.
\end{eqnarray}

\begin{figure}
	\includegraphics[scale=0.65]{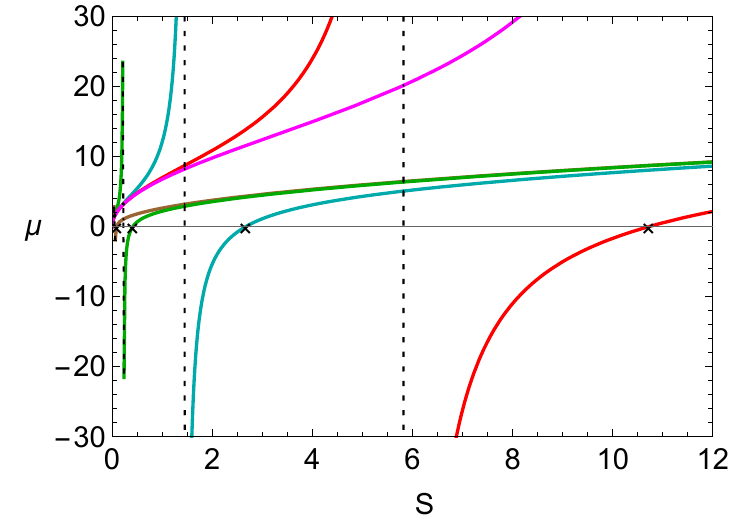}
	\hskip 1 cm
	\includegraphics[scale=0.65]{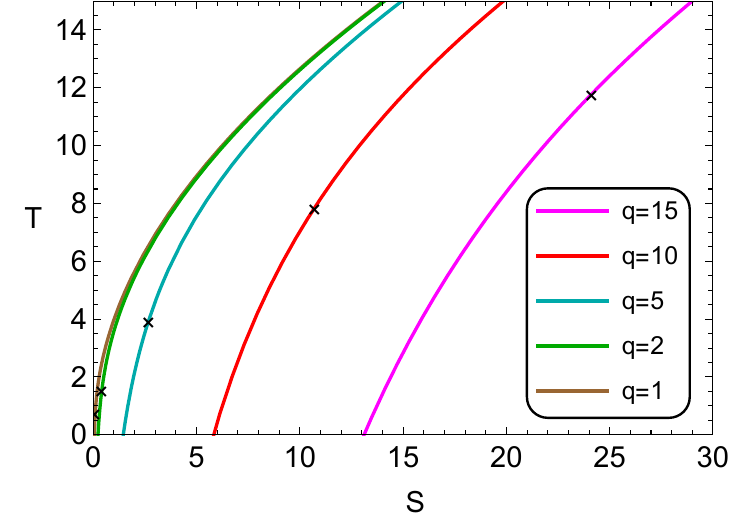}
	\caption{ \it{$\mu$ vs $S$ (left panel) and $T$ vs $S$ (right panel) 
for spherical charged BTZ-like black holes in $4$-dimensions, for fixed  $P=1$ 
and different values of $q$. The divergent point of the Joule-Thomson  
coefficient, which is denoted by the dashed vertical lines in the left panel, 
corresponds to the zero temperature in the right panel. There exist only minimum 
inversion temperatures, which are shown by cross marks both in the left and 
right panels, implying that cooling eventually occurs as the entropy 
increases.}}
	\label{musfig}
\end{figure}

The JT coefficient for spherical charged BTZ-like black holes in $4$-dimensions 
against entropy for the fixed value of pressure (as $P=1$) and different values 
of $q$ is depicted in the left panel of Fig. \ref{musfig}. The Hawking 
temperature versus entropy for the fixed values of $q$ and $P$ is also shown in 
the right panel of this figure. For each curve, the divergent point of the JT 
coefficient, which is denoted by the dashed vertical lines, corresponds to the 
zero temperature part of the corresponding curve in the right panel. In fact, 
for each $q$, the region to the left of the dashed vertical line is unphysical 
since the corresponding temperature takes negative values. It is observed that, 
for a fixed value of $q$, there exists only a minimum inversion temperature 
(shown by cross marks in the left and right panels), i.e., cooling occurs at 
large entropies. Setting $\mu=0$, the inversion pressure reads
\begin{equation}\label{pinv}
{P_i} = \frac{1}{\pi }\left[ {{2^{\frac{{D - 9}}{2}}}\left( {2D - 3} 
\right){q^{D - 1}}{{\left( {\frac{{4S}}{{{\Sigma _{(k)}}}}} \right)}^{\frac{{D 
- 1}}{{2 - D}}}} - {2^{\frac{{4(D - 1)}}{{D - 2}}}}k\,D\left( {D - 3} 
\right){{\left( {\frac{S}{{{\Sigma _{(k)}}}}} \right)}^{\frac{2}{{2 - D}}}}} 
\right].
\end{equation}
Using the Hawking temperature (\ref{temperature}), the inversion temperature 
can be expressed in terms of $P_i$ . For $k=0$, the inversion curve satisfies 
the following equation
\begin{equation}\label{k0}
{T_i} = \frac{{4\left( {D - 1} \right){\pi ^{\frac{1}{{1 - D}}}}}}{{\left( {D - 
2} \right)\left( {2D - 3} \right)}}q{\left[ {\frac{{{2^{\frac{1}{2}\left( {D - 
9} \right)}}\left( {2D - 3} \right)}}{{{P_i}}}} \right]^{\frac{1}{{D - 
1}}}}{P_i} \,.
\end{equation}
For $k=1$ and $-1$, it is not possible to attain a general equation for the 
inversion curve which explicitly expressed in terms of $D$ similar to what we 
found in eq. (\ref{k0}) for $k=0$. Instead one can derive an expression for the 
inversion equation for each specific $D$. Since the obtained equations are 
rather lengthy, we will display the corresponding inversion curves in the 
following figures. Letting $D=3$ in eqs. (\ref{pinv}) and (\ref{k0}), the 
inversion pressure and temperature for the charged BTZ black holes are given by
\begin{equation}
{\left. {{P_i}} \right|_{D = 3}} = \frac{{{\Sigma _{(k)}}3{q^2}}}{{128\pi 
{S^2}}} \quad \text{and} \quad
{\left. {{T_i}} \right|_{D = 3}} = 2\sqrt {\frac{2}{{3\pi }}} q{P_i}^{1/2},
\end{equation}
which exactly reproduces the results found in \cite{JT2021} by setting $\Sigma 
_{(k)}=2\pi$ and $q=Q/2$.  \vspace{1mm}

\begin{figure}
	\includegraphics[scale=0.65]{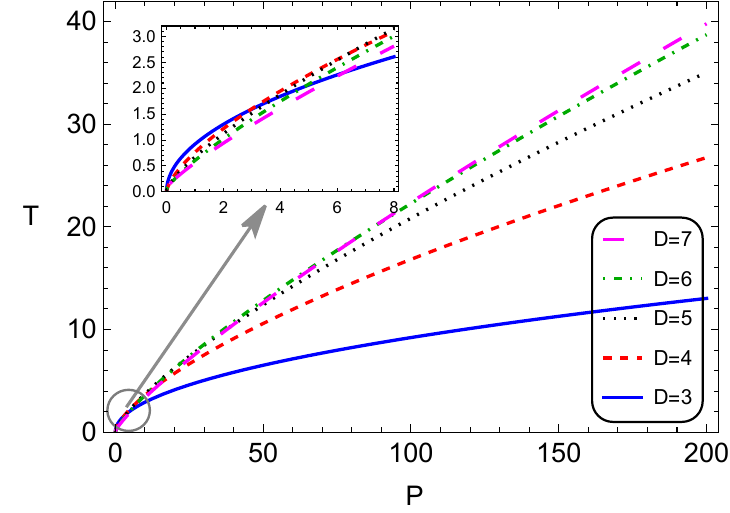}
	\hskip 1 cm
	\includegraphics[scale=0.65]{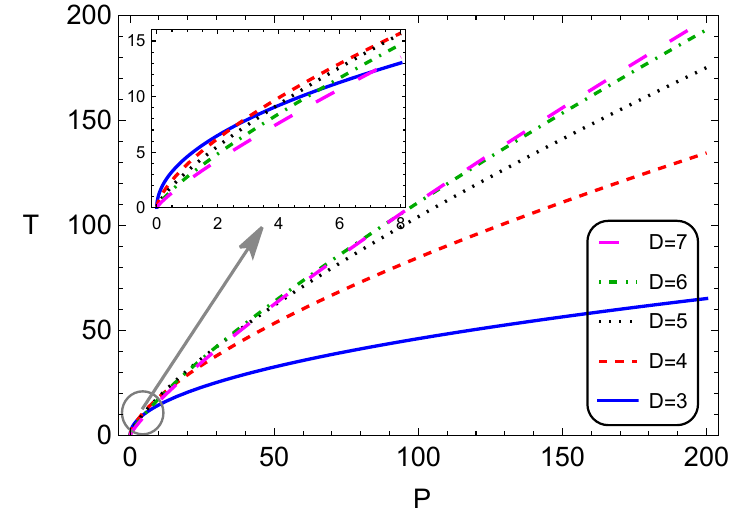}
	\caption{\it{The inversion curves of the conventional charged BTZ black 
hole ($D=3$), as well as of the spherical charged BTZ-like black holes in 
arbitrary dimensions ($D=4,5,6,7$), in scheme II, for $q=1$ (left 
panel) and $q=5$ (right panel). Cooling and heating regions lie above and below 
the inversion curves, respectively.}}
	\label{inv_k1_q15_d}
\end{figure}

\begin{figure}
	\includegraphics[scale=0.7]{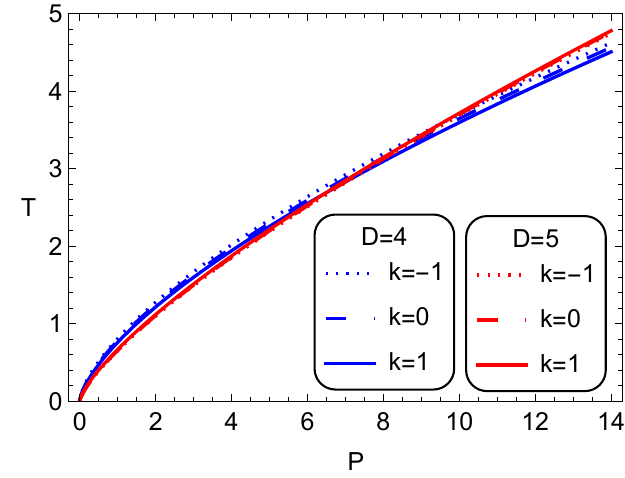}
	\caption{\it{The inversion curves of the spherical, planar and hyperbolic 
charged BTZ-like black holes in $D=4$ and $D=5$ in scheme II, for $q=1$. 
Cooling and heating regions lie above and below the inversion curves, 
respectively.}}
	\label{inv_d45_k}
\end{figure}

\begin{figure}
	\includegraphics[scale=0.46]{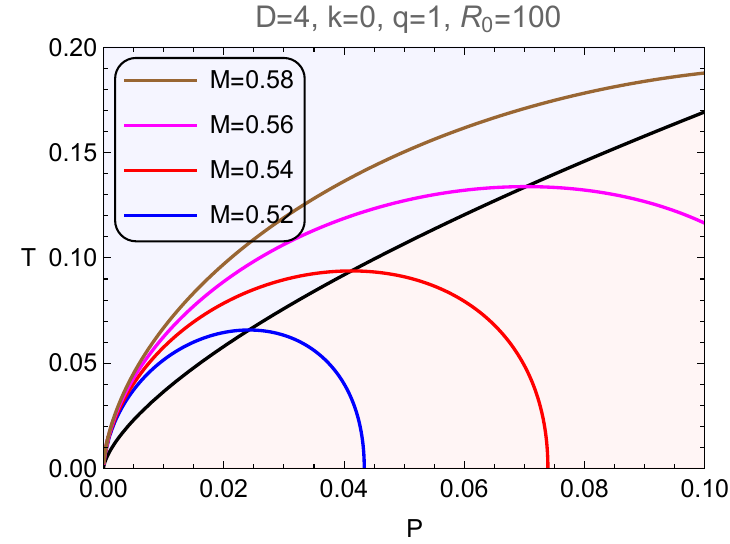}
	\hskip 0.1 cm
	\includegraphics[scale=0.47]{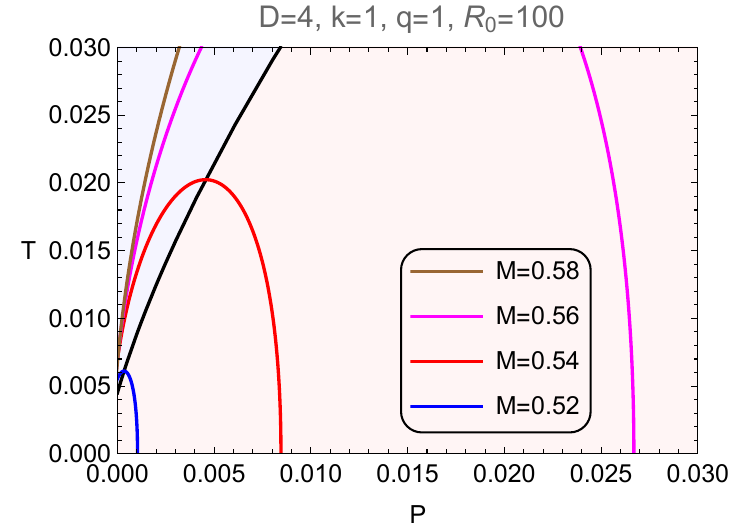}
	\hskip 0.1 cm
	\includegraphics[scale=0.46]{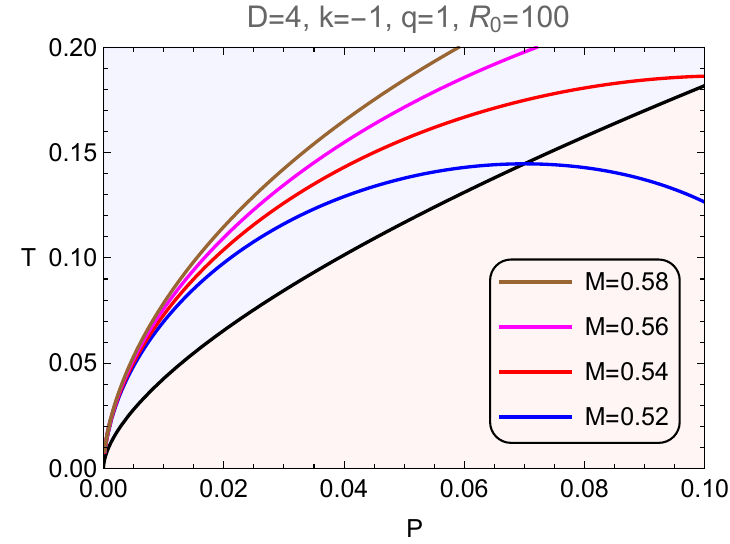}
	\caption{\it{The inversion curves (black curves) and the isenthalps 
(colored curves) of planar, spherical and hyperbolic charged BTZ-like black 
holes in $D = 4$, for   $q=1$ and $R_0=100$. For the blue/red regions, 
cooling/heating occurs as a result of decreasing pressure at constant 
enthalpy.}}
	\label{isen_curves}
\end{figure}

The inversion curves of the spherical charged BTZ-like black holes are plotted 
in spacetime dimensions $D=4,5,6$ and $7$ for two values of $q$ in Fig. 
\ref{inv_k1_q15_d}. Besides, the inversion curve for the conventional charged 
BTZ black hole in three-dimensions is depicted for comparison. The curves 
monotonically increase with pressure, i.e., there exist only minimum inversion 
temperatures and no maximum ones in conflict with real gases  for which the 
inversion curves are closed with both minimum and maximum inversion 
temperatures. This is a typical behavior of many AdS black holes for which the 
Joule-Thomson expansion is studied before 
\cite{JT2017,JT2018a,JT2018b,JT2019,JT2021}. The cooling and heating regions 
lie above and below the inversion curves, respectively. It is observed that, as 
the spacetime dimensionality increases from $D=3$ to $D=7$, the cooling region 
becomes smaller at high enough pressures. However, this is not the case at low 
enough pressures where the reverse behavior is seen; in such situations, the 
higher is the dimension the larger is the cooling region. A close-up of this 
behavior is depicted in the zoomed-in region of Fig. \ref{inv_k1_q15_d}. 
Finally, comparing the left and the right panels of Fig. \ref{inv_k1_q15_d} 
shows that, by decreasing the charge $q$, cooling occurs over a larger domain 
of the $T-P$ plane. In Fig. \ref{inv_d45_k}, the inversion curves for different 
geometries of the event horizon (i.e., $k=0, \pm 1$) are illustrated for $D=4$ 
and $D=5$. It can be seen that for a certain dimension, the inversion curves 
for different geometries are very close to each other. Setting $k=0$ and $D=4$ 
in eq. (\ref{thermo_schemeII}) and solving for $S$
\begin{equation}\label{isens}
S =  - \frac{{\sqrt[3]{{ - 1}}\,{q^2}}}{{8{\pi ^{2/3}}{P^{2/3}}}}{W_{ - 
1}}{\left\{ { - \frac{{2\sqrt 2 \pi PR_0^3}}{{{q^3}}}\exp \left[ { - 
\frac{{6\sqrt 2 \pi M}}{{{q^3}}}} \right]} \right\}^{2/3}},
\end{equation}
and putting this entropy into the temperature relation in 
(\ref{thermo_schemeII}), the isenthalpic curves in the $T-P$ plane are given by
\begin{equation}\label{isenc}
T = \frac{{{\Sigma _{(k)}}{{\left( { - 1} \right)}^{2/3}}\sqrt 2 
{P^{2/3}}\,q}}{{{\pi ^{1/3}}W}}\left\{ {{{\left[ {\frac{{{{\left( { - 1} 
\right)}^{4/3}}W}}{{{\Sigma _{(k)}}}}} \right]}^{3/2}} - 1} \right\}
\end{equation}
where
\begin{equation}
W = {W_{ - 1}}{\left\{ { - \frac{{2\sqrt 2 \pi 
PR_0^3}}{{{q^3}}}{\rm{exp}}\left[ { - \frac{{6\sqrt 2 \pi M}}{{{q^3}}}} \right]} 
\right\}^{2/3}}
\end{equation}
is the Lambert W function. The isenthalpic curves obtained in eq. (\ref{isenc}) 
are shown in the left panel of Fig. \ref{isen_curves} for $q=1$ and $R_0=100$ 
while setting ${\Sigma _{(k)}}=1$. The gradients of the isenthalps change sign 
as they cut the inversion curve. The region in which the isenthalps have 
positive (negative) slopes is the cooling (heating) region colored by blue 
(red) corresponding to $\mu>0$ ($\mu<0$). For $k=\pm 1$, analytic calculations 
similar to those leading to eqs. (\ref{isens}) and (\ref{isenc}) are not 
possible, so we performed numerical calculations in order to sketch the 
isenthalps in the middle and right panels of Fig. \ref{isen_curves}. \vspace{1mm}

In conclusion, the charged BTZ-like black holes undergo the Joule-Thomson 
expansion as expected due to the interacting nature of the system and nothing 
unusual was observed during this thermodynamic process which confirms this 
scheme results in consistent outcomes.

\section{Summary and conclusions} \label{conclusion} \label{sect8:conclusion}

In this work  we   investigated  the thermodynamics and  features of 
 higher-dimensional BTZ-like black-hole solutions,  namely higher-dimensional 
charged black holes in which   the electromagnetic sector exhibits the same 
properties with that of the usual  three-dimensional BTZ solution. In 
particular, we considered  
general relativity minimally coupled to Maxwell Lagrangian (${\cal F}$) in an 
(A)dS 
background, including  power of $(D-1)/2$ in order to extend 
the physics related to the electromagnetic sector of usual
BTZ black holes to higher dimensions. Assuming static ansatzes for both the 
gauge field ($A_\mu$) and the spacetime metric ($g_{\mu \nu}$), it was argued 
that the corresponding gauge field and black hole solutions of the 
electromagnetic and the gravitational field equations have to be as those 
presented in eqs. (\ref{gauge vector}) and (\ref{metric function}), 
respectively. After that, we evaluated the semi-classical partition function of 
the so-called charged BTZ-like black hole solutions (not calculated before) 
using the Euclidean path integral formalism and, by having that, the conserved 
charges were computed in agreement with other methods. The computation of the 
Euclidean action using the subtraction method in Sec. 
\ref{sect3:renormalization} revealed that inserting a renormalization length 
scale (which is a radial cutoff denoted by $R_0$) as a new integration constant 
in the metric function (\ref{metric function}) is essential to get finite 
results and thermodynamically obtain a consistent system. Interestingly, the $D 
\to 3$ limit of the computation of Euclidean on-shell action provides strong 
evidence that a renormalization length scale is required in the geometry of the 
conventional charged BTZ black holes. \vspace{1mm}

Black holes having a length cutoff in their geometry are thermodynamically 
challenging to address, especially when they are studied in the extended phase 
space thermodynamics, since such property leads to the emergence of different 
thermodynamic schemes. Since the theory entails a hard cutoff in the static 
black holes' geometry even in AdS background, naturally this question arises 
how to identify the length cutoff. This is interesting because it turns out 
that the black hole thermodynamics depends on identifying $R_0$ especially in 
the extended phase space. Also, the $D \to 3$ limit of the black hole solutions 
(reviewed in Sec. \ref{sect2:action}) coincides with the three-dimensional 
charged BTZ black holes, and additional degrees of freedom in higher dimensions 
enrich the thermodynamic phase space in comparison with the conventional 
charged BTZ black holes, which gives the study of this class of black holes a 
special significance. The traditional and somewhat problematic option is to 
identify the length cutoff with the AdS radius, $R_0=L$. Regarding this 
identification, a limit to asymptotically flat background of (A)dS black hole 
solutions (as usual, via $\Lambda \to 0$ or equivalently $L \to \infty$) is not 
possible. On the other hand, a gauge-invariant definition of the electric 
potential becomes ambitious. However, treating $R_0$ as an independent 
parameter leads to a well-defined gauge-invariant definition for the electric 
potential difference and also a limit to asymptotically flat background of (A)dS 
black hole solutions (\ref{metric function}) is accessible.  \vspace{1mm}

Next, we revisited the thermodynamics of charged BTZ-like black holes in 
arbitrary dimensions by
extending the thermodynamic phase space, i.e., treating the cosmological 
constant as a thermodynamic
variable, namely pressure ($\Lambda=-8 \pi P$), to see what insights can be 
gained from this modern context. It was found that there exist two possible 
thermodynamic schemes, for which the outcomes are drastically different and 
depend on how the length scale is identified. This can provide us further 
information about the role of the length cutoff in thermodynamics. We 
investigated the thermodynamic schemes and presented the first law as well as 
the corresponding Smarr relations. It was found that in the first scheme by 
identifying $R_0=L$, that we called scheme I, AdS black holes violate the 
conjectured RII, so they can be regarded as a new class of super-entropic black 
holes constructed in $D \ge 4$ dimensions. It was found that violating the RII 
conjecture (${\cal R} < 1$) in the first scheme (I) corresponds to the 
appearance of fundamental thermodynamic instabilities. There do not exist such 
thermodynamic instabilities whenever the RII conjecture is respected, ${\cal R} 
\ge 1$, as it happens in the second scheme in which the length cutoff is 
identified as an independent variable and consequently the RII conjecture is 
saturated (${\cal R} = 1$). In conclusion, the violation of the RII conjecture 
corresponds to the appearance of fundamental thermodynamic instabilities, 
confirming that the instability conjecture for this new family of 
super-entropic black holes is valid. \vspace{1mm}

For charged BTZ black holes as well as their generalization to higher 
dimensions, it was shown that
identifying the length cutoff as an independent thermodynamic variable leads to 
another kind of thermodynamic
instability in the second scheme since the energy of the system $E$ must be a 
convex function of its natural extensive variables including $R_0$, but this 
condition is not satisfied here, as shown in eq. (\ref{instability_R0}). This 
implies that it is mandatory to work in an ensemble where the length cutoff is 
kept fixed and is not allowed to fluctuate. Of course, thermodynamic 
instabilities can be eliminated if the length cutoff is treated as an 
independent and fixed parameter of the theory, but at the price of returning 
the inconsistency between the first law and the Smarr relation and, more 
importantly, such a scheme cannot be supported by the context of extended phase 
space thermodynamics.  \vspace{1mm}

We then focused on the other thermodynamic aspects of the AdS black hole 
solutions of this theory including the critical phenomena and the thermodynamic 
process of Joule-Thomson expansion. It was explicitly shown that the same 
behaviors for the isobars and the isotherms in the $T-r_+$ and the $P-r_+$ 
planes are found in both schemes which are quite analogous to the liquid/gas 
phase transition occurring in the vdW system. To our knowledge, charged 
BTZ-like black holes in scheme I are the first demonstration of super-entropic 
black holes which possess second-order critical points, so it was theoretically 
interesting to investigate the corresponding $P-v$ criticality and phase 
transition. The analyses of the Gibbs free energy in both schemes through the 
$G-T$ and the $G-P$ diagrams revealed that the corresponding vdW phase 
transition cannot be explained in the first scheme (because of the fundamental 
thermodynamic instabilities) while the second scheme provides a well-defined 
and consistent explanation in an ensemble where $R_0$ is kept fixed. For the 
sake of completeness, the process of Joule-Thomson expansion was also 
investigated in the second scheme as a consistency check. We investigated the 
Joule-Thomson coefficient, the inversion temperature, the inversion curves in 
the $T-P$ plane, and also the isenthalps, indicating that these class of black 
holes can be considered as interacting statistical systems.  \vspace{1mm}

This study suggests that the traditional treatment of charged BTZ black hole 
thermodynamics as well as its generalizations via identifying the length cutoff 
($R_0$) with the AdS radius ($L$), the so-called thermodynamic scheme I, is 
probably not fully consistent (following \cite{Mo2017,Appels2020,JT2021}) or, 
at least, it leads to a number of unusual, pathological thermodynamic 
implications due to the fundamental thermodynamic instabilities. As seen, more 
insights about black hole systems can be gained via the extended phase space 
thermodynamics as a diagnosis, especially when different schemes come to play. 
On the other hand, since many examples of nonlinearly charged BTZ black holes 
violating the RII conjecture can be constructed by coupling gravity to nonlinear 
$U(1)$ gauge-invariant modifications of Maxwell's classical electrodynamics in 
three-dimensions, we speculate that the results of this research will be valid 
for such cases as well if the weak-field limit of them matches that of 
Euler-Heisenberg electrodynamics (see some reviews of nonlinear theories of 
electrodynamics in refs. \cite{DSZ2022,Sorokin2022}); There will be two 
different thermodynamic schemes and, in the thermodynamic scheme where $R_0 = 
L$, AdS black holes are super-entropic and suffer from fundamental 
thermodynamic instabilities. While, in the other scheme where $R_0 \ne L$, we 
expect the same behaviors as found here to show up qualitatively. However, it 
is interesting to investigate whether there exists thermodynamic instability 
associated with the length cutoff in such theories or not.   \vspace{1mm}

There are still some potential avenues to further study the charged BTZ-like 
black holes or those black hole spacetimes from coupling gravity in 
three-dimensions to nonlinear electrodynamics within the context of extended 
phase space thermodynamics in future, which can improve our understanding of 
black holes having a cutoff in their geometry. Especially, it is important to 
investigate the thermodynamic properties of such black holes in the first 
scheme (I), where they enjoy super-entropicity, to seek whether there exists a 
universal behavior near the critical point or not. Considering such black hole 
solutions as a working substance of a new class of holographic heat engines is 
another interesting task. Investigating microscopic structure and quantum 
corrections to the various thermodynamic schemes have also their own special 
importance. One more crucial subject to consider is the holographic dual interpretation of extended black hole thermodynamics across different thermodynamic schemes. Another interesting subject is the geometric and thermodynamic properties of the obtained black holes in dS space. This is interesting since we confirmed that, unlike the AdS case, charged BTZ-like black holes with dS asymptote are not super-entropic in any scheme and they always respect the RII conjecture. Cosmological applications of the obtained black holes in dS space are of interest since the length (radial) cutoff in the black hole geometry acts as a (thermal) cavity wall, leading to a thermal equilibrium between the event and cosmological horizons. This property makes them interesting to further investigate in future. In this regard, explorations into the difference between the thermodynamic schemes in dS space could be instructive as well. We will address some of these issues in future works. \vspace{1cm}

{\bf{\ \ \ \ \ \ \ \ \ \ \ \ \ \ \ \ \ \ \ \ \ \ \ \ \ \ \ \ \ \ \ \ \ \ 
\ \ \ \ \ \ \ \ \ \ \ \ \ \ \ \ \ \ \ \ \ \ 
 Acknowledgments}}\\

A.D. wish to thank the Iran Science Elites Federation and Damghan 
University. S.Z. wish to thank the  University of Sistan and 
Baluchestan's Research Council.


\end{document}